\documentclass[onecolumn]{aa}
\usepackage{graphicx}
\usepackage{txfonts}
\begin{document}
\newcommand{\magcir}{\ \raise -2.truept\hbox{\rlap{\hbox{$\sim$}}\raise5.truept
 	\hbox{$>$}\ }}	
\newcommand{\mincir}{\ \raise -2.truept\hbox{\rlap{\hbox{$\sim$}}\raise5.truept
 \hbox{$<$}\ }} 

  \title{The Great Observatories Origins Deep Survey}
  \subtitle{VLT/FORS2 Spectroscopy in the GOODS-South Field}  

  \author{E. Vanzella\inst{1,2}
     \and
      S. Cristiani\inst{2}
     \and 
      M. Dickinson\inst{3}
     \and 
      H. Kuntschner\inst{4}
     \and
      L. A. Moustakas\inst{5}
     \and 
      M. Nonino\inst{2}
     \and 
      P. Rosati\inst{6}
     \and 
      D. Stern\inst{8}
     \and
      \\
      C. Cesarsky\inst{6}
     \and 
      S. Ettori\inst{6}
     \and
      H. C. Ferguson\inst{5}
     \and 
      R.A.E. Fosbury\inst{4}
     \and
      M. Giavalisco\inst{5}
     \and
      J. Haase\inst{4}
     \and 
      A. Renzini\inst{6}
     \and
      A. Rettura\inst{6,7}
     \and
      \\
      P. Serra\inst{4}
     \and
     the GOODS Team 
     }

  \offprints{E. Vanzella (evanzell@eso.org)}

  \institute{Dipartimento di Astronomia dell'Universit\`a di Padova,
    Vicolo dell'Osservatorio 2,
    I-35122 Padova, Italy.  
     \and
       INAF - Osservatorio Astronomico di Trieste, Via G.B. Tiepolo 11,
       40131 Trieste, Italy.
     \and 
       National Optical Astronomy Obs., P.O. Box 26732, Tucson, AZ 85726.
     \and 
       ST-ECF, Karl-Schwarzschild Str. 2, 85748 Garching, Germany.
     \and
       Space Telescope Science Institute, 3700 San Martin Drive,
       Baltimore, MD 21218. 
     \and
       European Southern Observatory, Karl-Schwarzschild-Strasse 2,
Garching, D-85748, Germany.
     \and 
     Universite' Paris-Sud 11, Rue Georges Clemenceau 15, Orsay, F-91405, France
     \and
     Jet Propulsion Laboratory, California Institute of Technology,
     MS 169-506, 4800 Oak Grove Drive, Pasadena, CA 91109
     \thanks{Based on observations made at the European Southern
Observatory, Paranal, Chile (ESO programme 170.A-0788 {\it The Great
Observatories Origins Deep Survey: ESO Public Observations of the
SIRTF Legacy/HST Treasury/Chandra Deep Field South.}) }
       }
  \date{Received ; accepted }

 \abstract{
 We present the first results of the ESO/GOODS program of spectroscopy
 of faint galaxies in the Chandra Deep Field South (CDF-S). 399 spectra 
 of 303 unique targets have been obtained in service mode with the FORS2
 spectrograph at the ESO/VLT, providing 234 redshift
 determinations (the median of the redshift distribution is at 1.04). 
 The typical redshift uncertainty is estimated to be
 $\sigma_z \simeq 0.001$. Galaxies have been color selected in a way
 that the resulting redshift distribution typically spans from z=0.5 to 2.
 The reduced spectra and the derived redshifts are released to the community 
 through the ESO web page $\it{http://www.eso.org/science/goods/}$. 
 Large scale structure is clearly detected at $z \simeq 0.67, 0.73, 
 1.10$ and $1.61$. Three Lyman-break galaxies have also been included as
 targets and are confirmed to have redshifts $z=4.800$, $4.882$ and $5.828$.
 In a few cases, we observe clear [OII]3727 rotation curves,
 even at the relatively low resolution ($\Re = 860$) of the present observations. 
 Assuming that the observed velocity structure is due to dynamically-relaxed
 rotation, this is an indication of large galactic masses (few times $10 ^{11} M_\odot$)
 at $z \sim 1$.

 }

 \maketitle
%

\section{Introduction}
The Great Observatories Origins Deep Survey (GOODS) is a public,
multifacility project that aims to answer some of the most profound
questions in cosmology: how did galaxies form and assemble their
stellar mass? When was the morphological differentiation of galaxies
established and how did the Hubble Sequence form? How did AGN form and
evolve, and what role do they play in galaxy evolution? How much do
galaxies and AGN contribute to the extragalactic background light? Is
the expansion of the universe dominated by a cosmological constant? A
project of this scope requires large and coordinated efforts from many
facilities, pushed to their limits, to collect a database of
sufficient quality and size for the task at hand. It also requires
that the data be readily available to the worldwide community for
independent analysis, verification, and follow-up.

The program targets
two carefully selected fields, the Hubble Deep Field North (HDF-N) and
the Chandra Deep Field South (CDF-S), with three NASA Great
Observatories (HST, Spitzer and Chandra), ESA's XMM-Newton, and a wide
variety of ground-based facilities. The area common to
all the observing programs is 320 arcmin$^2$, equally divided between
the North and South fields. For an overview of GOODS, see \cite{dick03}, 
\cite{renz02} and \cite{giava04a}. 

Spectroscopy is essential to reach the scientific goals of GOODS.
Reliable redshifts provide the time coordinate needed 
to delineate the
evolution of galaxy masses, morphologies, clustering, and star 
formation. They calibrate the photometric redshifts that can 
be derived from the imaging data at 0.36-8$\mu$m. 
Spectroscopy will measure physical
diagnostics for galaxies in the GOODS field (e.g., emission line
strengths and ratios to trace star formation, AGN activity,
ionization, and chemical abundance; absorption lines and break
amplitudes that are related to the stellar population ages). Precise redshifts are
also indispensable to properly plan for future follow-up at higher
dispersion, e.g., to study galaxy kinematics or detailed spectral-line
properties.

The ESO/GOODS spectroscopic program is designed to observe
all galaxies for which VLT optical spectroscopy is likely to yield
useful data.
The program makes full use of the VLT instrument capabilities (FORS2 and VIMOS),
matching targets to instrument and disperser combinations in order to
maximize the effectiveness of the observations. The magnitude limits
and selection bandpasses depend to some degree on the instrumental
setup being used. The aim is to reach mag~$\sim24-25$ with adequate S/N, with
this limiting magnitude being in the B band for objects observed with
the VIMOS LR-Blue grism, in the V band for those observed in the VIMOS
LR-Red grism, and in the z band for the objects observed
with FORS2. 
This is not only a practical limit, however, but is also
well matched to the scientific aims of the GOODS program. The ACS
$i_{775}$ imaging samples rest-frame optical (B-band) light out to $z
= 1$, where $i_{775} = 25$ reaches 1.5 to 2 magnitudes past $L^{\ast}_B$. 
This is also the practical limit for high-quality, quantitative
morphological measurements from the ACS images (cf. \cite{abraham96}).
Similarly, $i_{775} = 25$ is $\sim 1$ mag fainter than the
measured $L^{\ast}$ UV for $z = 3$ Lyman Break Galaxies (LBGs), 
and 0.5 mag fainter than that at $z = 4$ (\cite{steidel99}). 
These are the limits to which
GOODS/SIRTF IRAC data will robustly measure rest-frame near-IR light,
and hence constrain the stellar mass.

In this paper we report on the first spectroscopic follow-up campaign in the
Chandra Deep Field South (CDF-S), carried out with the FORS2
instrument at the ESO VLT in the period fall 2002 - spring 2003
(the first 9 masks, 348 slits). Further 17 masks have been observed during
the period 2003 and early 2004, for which the reduction process has
started and will be presented elsewhere (Vanzella et al., in 
preparation).

The paper is organized as follows: in Sect. 2 we describe the target selection 
and in Sect. 3 the observations and the reduction. 
The redshift determination is presented in Sect. 4. In Sect. 5 we discuss
the data and in Sect. 6 the conclusions are presented. 
Throughout this paper the magnitudes are given in the AB system 
(AB~$\equiv 31.4 - 2.5\log\langle f_\nu / \mathrm{nJy} \rangle$),
and the ACS F435W, F606W, F775W, and F850LP filters are denoted 
hereafter as $B_{435}$, $V_{606}$, $i_{775}$ and $z_{850}$, respectively. 
We assume a cosmology with $\Omega_{\rm tot}, \Omega_M, \Omega_\Lambda = 1.0, 0.3, 0.7$
and $H_0 = 70$~km~s$^{-1}$~Mpc$^{-1}$.


\section{Target Selection}

Objects were selected as candidates for FORS2 observations
primarily based on the expectation that the detection and measurement
of their spectral features would benefit from the high throughput and
spectral resolution of FORS2, and its reduced fringing at red
wavelengths,
relative to other instrumental options such as VIMOS.   In particular,
we expect that the main spectral emission and absorption features for
galaxies
at  $0.8 < z < 1.6$ would appear at very red optical wavelengths,
out to $\sim 1 \mu$m.  Similarly, very faint Lyman break galaxies
at $z \gtrsim 4$, selected as $B_{435}$, $V_{606}$ and
$i_{775}$--dropouts
from the GOODS ACS photometry, also benefit greatly from the
red throughput and higher spectral resolution of FORS2.

In practice, several categories of object selection criteria were used
to
ensure a sufficiently high density of target candidates on the sky to
efficiently fill out multi-slit masks.   Using ACS photometry in the AB
magnitude system, these criteria were:

\begin{enumerate}

\item{Primary catalog:  $(i_{775}-z_{850}) > 0.6$ and $z_{850} < 24.5$.
This should ensure redshifts $z \gtrsim 0.7$ for ordinary early-type
galaxies
(whose strongest features are expected to be absorption lines),
and higher redshifts for intrinsically bluer galaxies likely to have
emission lines.}

\item{Secondary catalog:  $0.45 < (i_{775}-z_{850}) < 0.6$ and $z_{850}
< 24.5$.}

\item{Photometric-redshift sample:  $1 < z_{phot} < 2$ and $z_{850} <
24.5$, using
an early version of GOODS photometric redshifts like those described by
\cite{moba04}.}

\item{$i_{775}$-dropout and $V_{606}$--dropout Lyman break galaxy
candidates, selected from the criteria of \cite{dick04} 
and \cite{giava04b}, respectively.}

\item{A few miscellaneous objects, including host galaxies of
supernovae detected in the GOODS ACS observing campaign.}

\end{enumerate}

Target selection and mask design for the 2002-3 GOODS/FORS2 campaign
was carried out while the GOODS ACS observations were still in progress,
and before final ACS data reduction or cataloging could be completed.
The targets were therefore selected based on interim data reductions
and catalogs, initially based on only one epoch of ACS imaging, and
later
using the three-epoch ACS stacks and preliminary catalogs described in
\cite{giava04a}.   Because of this, the actual magnitudes and
colors of the observed galaxies from the final, 5-epoch ACS image stack,
which we report here in Table 2, may not exactly match the intent of the
original selection criteria.   When designing the masks, we generally
tried to avoid observing targets that had already been observed in other
redshift surveys of this field, namely, the K20 survey of \cite{cimatti02} 
and the survey of X-ray sources by \cite{szo04}.

In the present spectroscopic catalog there are 303 targets,
114 meeting the primary selection criterion and 56 meeting the 
secondary selection criteria. The other targets belong to the 
remaining classes.

\section{Observations and Data Reduction}
The VLT/FORS2 spectroscopic observations were carried out 
in service mode during several nights in 2002 and 2003. 
A summary is presented in Table \ref{tab:tblobs}.
In all cases the $300$I grism was used as dispersing element without 
order-separating filter.
This grism provides a scale of roughly 3.2\AA/pixel. The nominal
resolution of the configuration was
$\Re =\lambda/\Delta\lambda$=860, which corresponds to about
9{\AA} at 8000{\AA}. The spatial scale of FORS2 was $0.126\arcsec$/pixel, 
the slit width was always $1\arcsec$.
Dithering of the targets along the slits was applied
in order to effectively improve the sky subtraction an the removal
of CCD cosmetic defects.
The mean shift applied was $\pm8$ pixels.
\begin{table}
\centering \caption{Journal of the MXU Observations.}
\begin{tabular}{lccc}
\hline \hline
 Mask ID & UT date & exp.time (s)\\
\hline
 990247 &30Dec.2002 - 2,6Jan. 2003&12$\times$1200\cr
 984829 &9Dec.2002 - 3,4Jan. 2003 & 12$\times$1200 \cr
 985831 &5Jan. - 4,7Feb. 2003 & 15$\times$1200 + 663 \cr
 973934 &7, 30, 31Jan. 2003&12$\times$1200 \cr
 952426 &6,7Jan. 2003 & 12$\times$1200  \cr
 981451 &31Jan. - 24,27Feb. - 22Nov. - 17Dec. - 30Jul. - 1Aug. 2003 & 24$\times$1200 \cr
 995131 &5-6 Oct. 2002 & 8$\times$1800 \cr 
 994852 &4 Oct. 2002 & 8$\times$1800 \cr
 990652 &8Dec. 12Nov. 2002 & 14$\times$1200 + 300 + 900\cr
\hline
\label{tab:tblobs}
\end{tabular}
\end{table}

\subsection{\it Data Reduction}

\begin{figure}
 \centering
 \includegraphics[width=\textwidth]{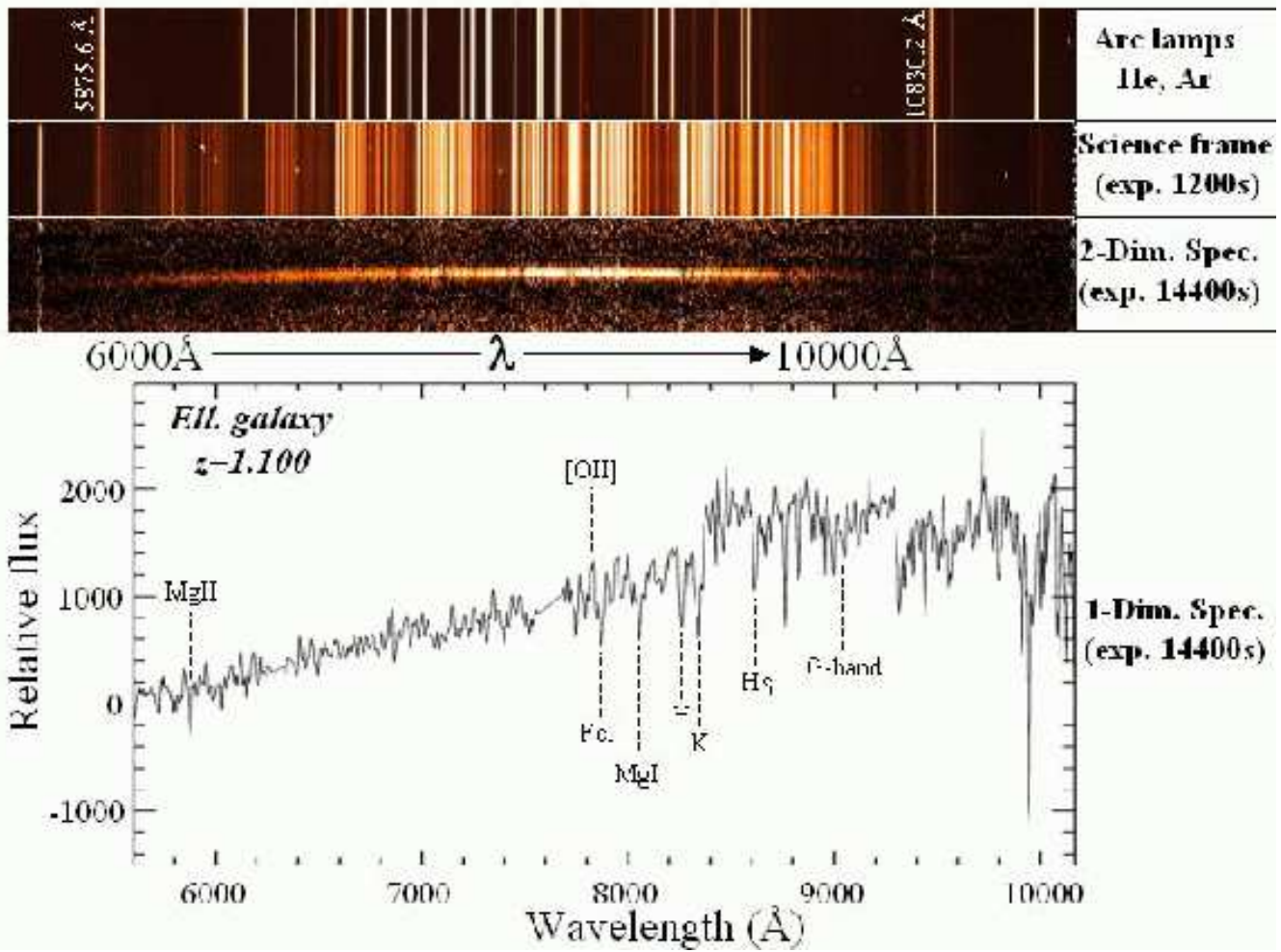}
 \caption{Typical FORS2 data products for an individual slit of multi-object mask.
 From the top of the figure: the 2-D spectrum of the arc lines used for the 
 wavelength calibration, a 2-D science exposure (1200 seconds), 
 the final flat-fielded and sky-subtracted 2-D spectrum (co-addition of 12 
 exposures for a total of 4 h), and at the bottom the 1-D spectrum with
 the identification of the main absorption and emission lines 
 (in this example an elliptical
 galaxy at z=1.100, $GDS~J$033217.46-275234.8).}
\label{fig:reduction}
\end{figure}

Data were reduced with a semi-automatic pipeline 
that we have developed on the basis of the MIDAS package (\cite{eso_midas},
using commands of the LONG and MOS contexts (Fig.~\ref{fig:reduction}). 
The frames have been bias-subtracted and flat-fielded.
For each slit the sky background was estimated
with a second order polynomial fitting. 
In some cases, better 
results have been obtained adopting a first order polynomial.
The fit has been computed independently in each column inside two windows, 
above and below the position of the object (if more objects are present 
in the slit, a suitable modification of the windows is applied). 

The resulting dithered, sky-subtracted, two-dimensional frames for each
object are then averaged, with the weighting determined based on exposure
time, seeing, and meteorlogical conditions.
Spatial median filtering has been applied to each dithered exposures to clean 
the cosmic rays. The FORS2 instrument shows an exquisite
response in the red domain (beyond 8000\AA), in practice no appreciable 
residual fringe pattern affects the extracted signal. 
Rather, the sky residuals dominate the noise in the regions where the 
intensity and the density of the skylines increase 
(see Figure~\ref{fig:reduction}).
The individual dithered sky-subtracted spectra have been visually
inspected to verify that the object is indeed in the expected region
of the slit.
This step is necessary since the applied small spatial offsets between
the science exposures can result in objects falling too close to the
slit edge or even outside the slit (in exceptional cases).
After this visual screening, the {\em spatial} offset between
different exposures of the same object was calculated on the basis of 
the {\em world coordinate system} (WCS) information stored in
the frame headers. 
The individual exposures were co-added (including the rejection of bad
pixels or cosmic ray hits) after applying these spatial shifts. The
frames were shifted in the spatial direction and only by integer
numbers of pixels.
As the objects were sufficiently well sampled (the pixel
scale was significantly smaller than the seeing), no significant
blurring of the spectra was observed, while the statistical 
properties of the individual pixels were preserved.

The position of the target on the detector was estimated by collapsing
700 columns in the dispersion direction and measuring the center of
the resulting profile.
The 1-D object signal was obtained using the `optimal extraction' method of 
MIDAS. This procedure calculates a weighted average in each column,
based on both the estimated object profile and photon
statistics.

Wavelength calibration was calculated on (daytime) arc
calibration frames, using three arc lamps (a He, and two Ar lamps)
providing sharp emission lines over the whole spectral
range used (6000--10800\AA).
The object spectra were then rebinned to a linear wavelength scale.
We have verified the accuracy of the wavelength calibration 
by checking the position of 25 narrow skylines 
in the science exposures (from 5577 to 10400 \AA).
Systematic translations of the wavelength scale have been typically measured 
to be of the order of $\pm$1\AA\ and corrected. The final (absolute)
mean wavelength accuracy for all spectra is
0.9$\pm$0.1\AA\ RMS.

Relative flux calibration was achieved by observations of standard
stars listed by \cite{bohlin95}. 
Since the standard stars are typically quite blue and second order
light can be substantial, the calibration spectra were obtained both
with and without order-sorting filters, providing calibration across 
the entire optical window. As noted previously, we opted to obtain 
the science target {\em without} an order-sorting filter, implying
deleterious effects to the flux calibration, particularly for
bluer objects and at longer wavelengths. For the red objects which
dominated the FORS2 target selection, we felt that the improved 
wavelength coverage more than compensated for the slightly 
comprimised flux calibration. Due to both this second order light
and uncertain slit losses, we caution against using the calibrated
fluxes for scientific purposes.

\section{Redshift Determination}
Spectra of 399 objects have been extracted. From them we have been
able to determine 234 redshifts. 
In the large majority of the cases the redshift has been determined through the
identification of prominent features of galaxy spectra:
the 4000\AA\ break, Ca H and K, g-band, MgII 2798, AlII 3584 in absorption and
$Ly_{\alpha}$, [O\,{\sc ii}]3727, [O\,{\sc iii}]5007, H$\beta$, H$\alpha$ 
in emission.
The redshift estimation has been performed cross-correlating the observed
spectrum with templates of different spectral types (S0, Sa, Sb, Sc, Ell.,
Lyman Break, etc.), using the $rvsao$ package in the IRAF environment.
The redshift identifications are summarized in Table~\ref{tab:tblspec} and
are available at the URL $\it{http://www.eso.org/science/goods/}$.

In Table~\ref{tab:tblspec}, 
the column {\em ID} contains the target identifier, that is constructed out of the 
target position (e.g., $GDS~J$033206.44-274728.8) where GDS stands for {\bf G}OO{\bf D}S {\bf S}outh. 
The {\em quality} flag, indicates the reliability of the
redshift determination. Quality ``A'' indicates a solid redshift determination,
``B'' a likely redshift determination, ``C'' a tentative redshift determination
and ``X'' an inconclusive spectrum or three cases in which no extraction was possible. 
150 objects have been classified with quality ``A'', 57 with quality ``B'',
27 with quality ``C'', 69 with inconclusive redshift determination ``X''.

\begin{figure}
 \centering
 \includegraphics[width=\textwidth,height=13cm]{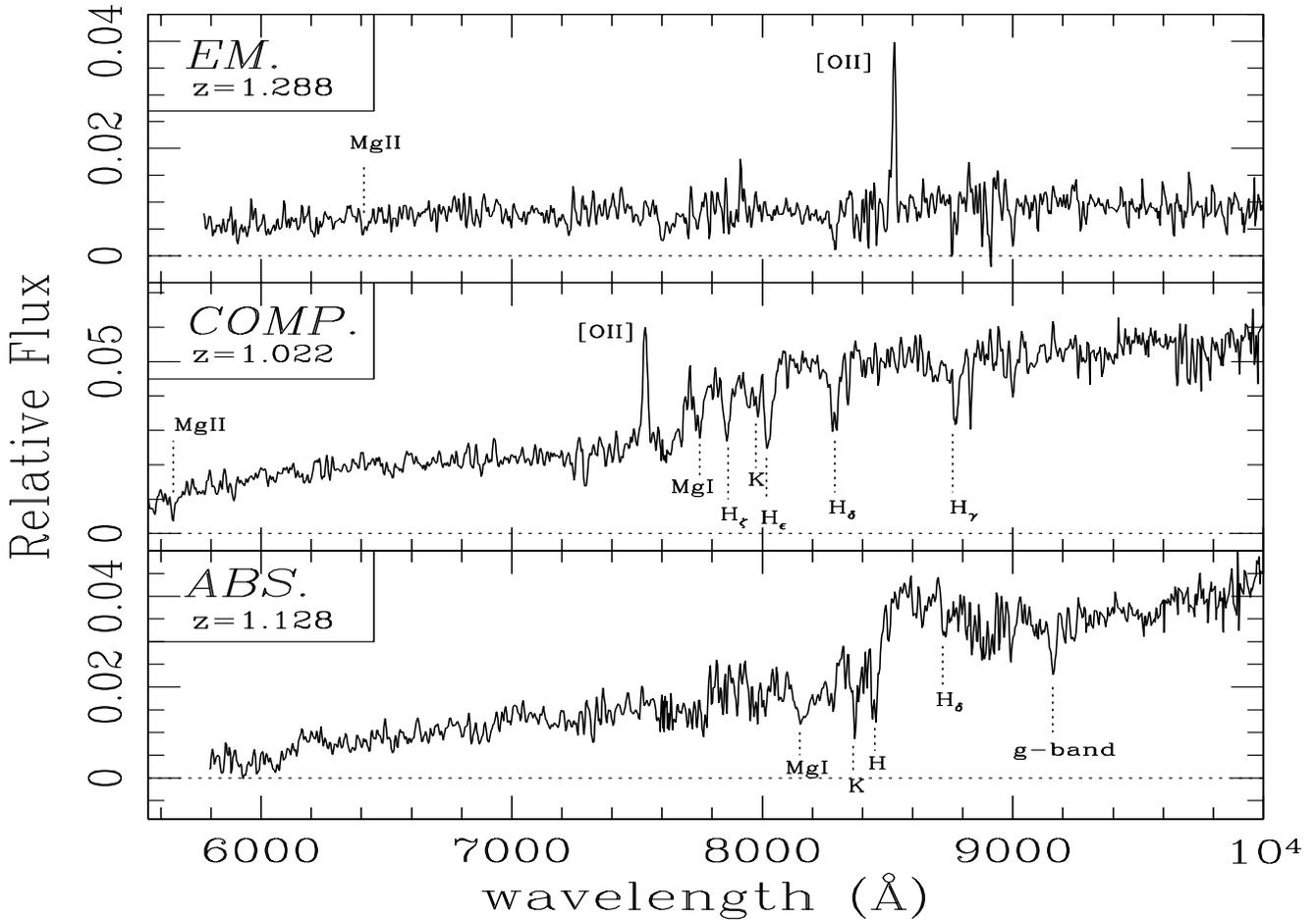}
 \caption{Three examples of objects classified as ``em.'' (emission-lines detected), 
``abs.'' (absorption lines) and 
``comp.'' (both emission and absorption lines detected).}
\label{fig:example_class}
\end{figure}

The {\em class} flag groups the objects for which emission line(s) (em.),
absorption-line(s) (abs.) or both (comp.) are detected in the spectrum.
The classification has been guided by the observed continuum level and
slope blueward and redward of the emission/absorption feature, by the 
broad-band colors and the morphology of the targets (see 
Figure~\ref{fig:example_class}). 11 objects have been classified as stars.

In 38$\%$ of the cases the redshift is based only on one emission line, usually
identified with [O\,{\sc ii}]3727 or $Ly_{\alpha}$. 
In these cases the continuum shape, the presence of breaks, the
absence of other spectral features in the observed spectral range and the broad band 
photometry are particularly important in the evaluation.
In general these solo-emission line redshifts are classified as ``likely'' (B)
or ``tentative'' (C). 

\begin{figure}
 \centering
 \includegraphics[width=\textwidth]{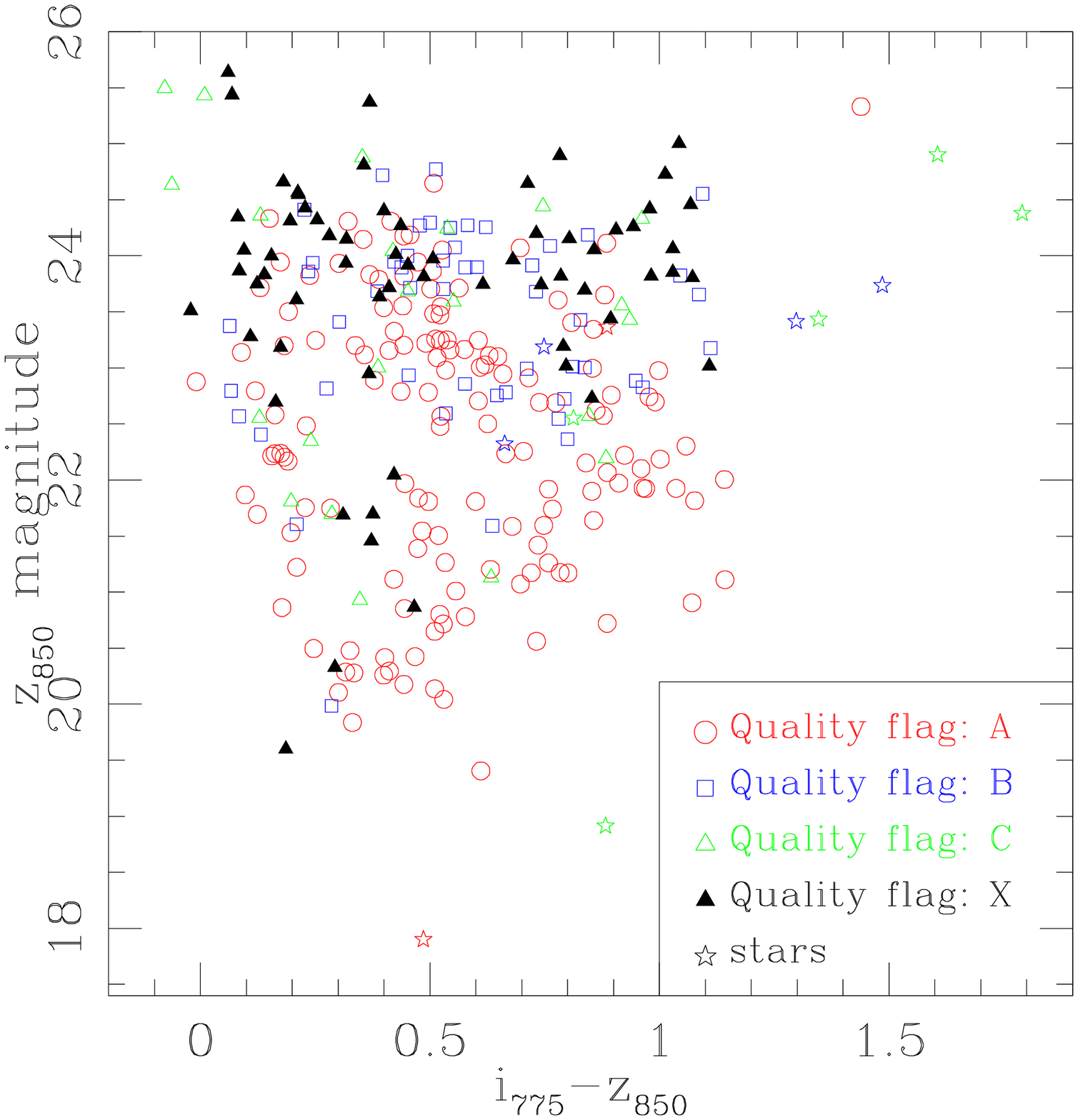}
 \caption{Color-magnitude diagram for the spectroscopic sample as a function of
the quality flag. The uncertainties in the redshift determination increase 
with increasing $z_{850}$ magnitude. Few bright sources (often serendipitous; 
$z_{850}<22$) have inconclusive redshift determinations due to the dithering 
procedure, which has positioned these sources off the slitlets for many of the exposures.}
\label{fig:zVSiz}
\end{figure}

Finally, the {\em comments} column contains additional information relevant
to the particular observation. The most common ones summarize 
the identification of the principal lines, the inclination of an emission 
line due to internal kinematics, the weakness of the signal 
(``faint''), the low S/N of the extracted spectrum (``noisy''), 
the $20\%$ light radius  (``Flux-radius'') for objects classified as stars, 
etc.

There are two objects that are not present in the v1.0 catalog, with
$z=0.957$ and $z=4.882$ (marked with a cross in the Table~\ref{tab:tblspec}).
These two objects were not successfully deblended in the detection process
from the brighter nearby galaxies.

\begin{figure}
 \centering
 \includegraphics[width=\textwidth]{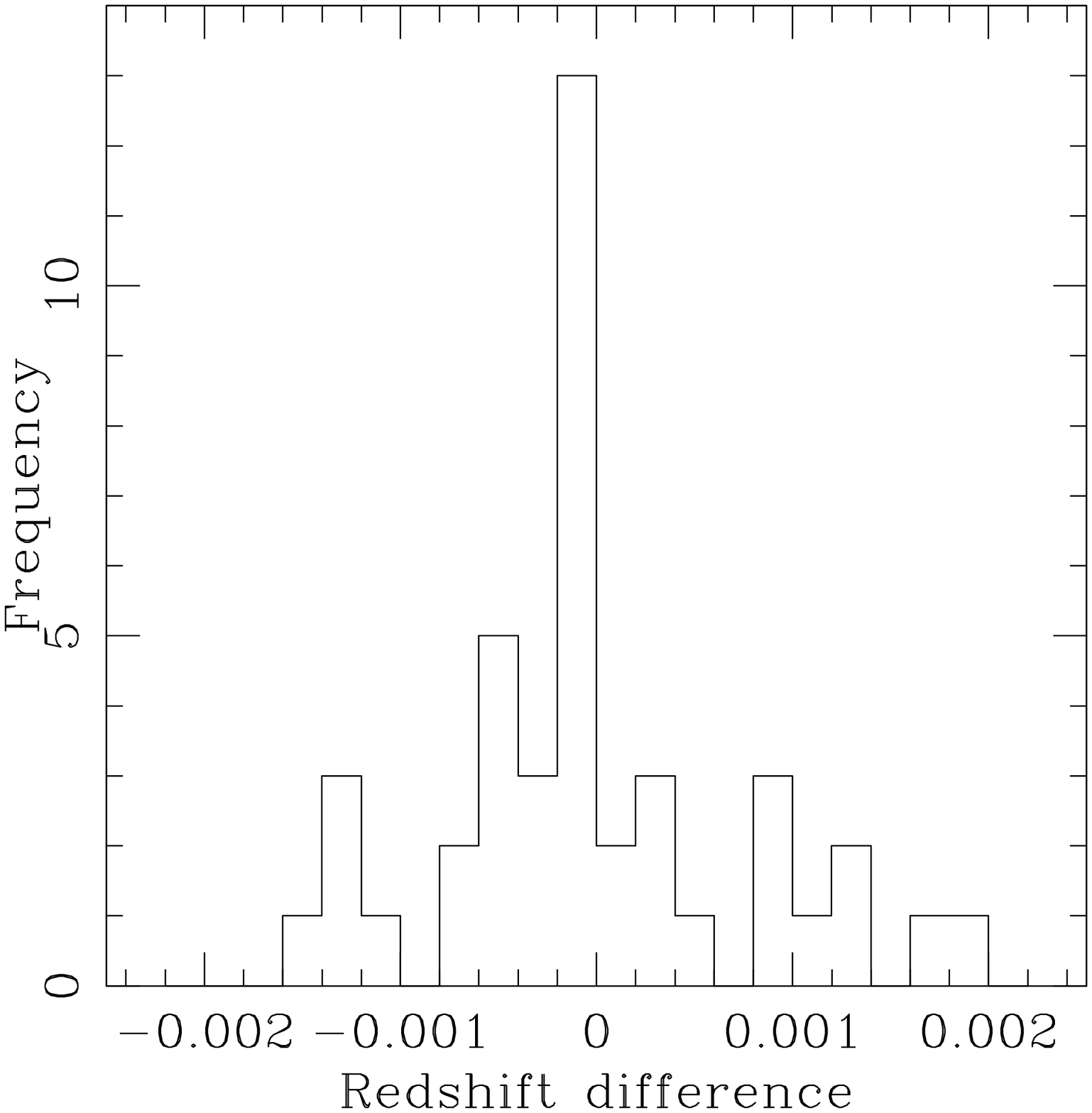}
 \caption{Redshift differences between objects observed twice or more in 
independent FORS2 observations. The distribution has a dispersion 
$\sigma_{z}=0.00078$.}
\label{fig:z_err}
\end{figure}

The internal redshift accuracy can be estimated from a sample of 42 galaxies 
which have been observed twice (or more) in independent FORS2 mask sets. 
The distribution of measured redshift differences is presented in 
Figure~\ref{fig:z_err}. The mean of the distribution is close to zero ($10^{-6}$)
and the redshift dispersion $\sigma_z=0.00078$ and mean absolute deviation
$< | \Delta z | > = 0.00055$, fairly constant with redshift.
These values can be considered as a lower limit to the redshift uncertainty.

\begin{table}
\caption{Spectroscopic redshift catalog.}
\begin{tabular}{lcccccl}
\hline \hline
  ID(V1)       & $z_{850}$      & $(i_{775}-z_{850})$  & zspec         &class. & Quality & comments \\
\hline
GDS~J033206.44-274728.8	&  21.07	&  0.70	&  1.022	&comp.	&A	& [OII], CaHK, MgI	 \cr 
GDS~J033210.73-274819.4	&  24.33	&  0.96	&  1.396 	&em.	&C	&faint, [OII]	 \cr 
GDS~J033210.79-274719.8	&  25.64	&  0.06	& - 	&-	&X	&faint(line@8069A?) 	 \cr 
GDS~J033210.92-274722.8	&  19.98	&  0.29	&  0.417	&em.	&B	&CaH, g-band, [OIII], Na, H$\alpha$ \cr 
GDS~J033210.93-274721.5	&  22.19	&  1.00	&  1.222	&abs.	&A	&CaHK, MgI, H$\delta$ \cr 
GDS~J033212.00-275104.2	&  23.00	&  0.85	&  1.018	&comp.	&A	&[OII], CaHK, g-band	 \cr 
GDS~J033212.47-274621.4	&  24.06	&  1.03	& - 	&-	&X	&faint	 \cr 
GDS~J033212.61-274605.1	&  24.05	&  0.53	&  1.378	&em.	&A	& [OII], MgI	 \cr 
GDS~J033212.79-274823.1	&  23.03	&  0.45	&  1.316	&em.	&A	& [OII], MgII, (CaHK faint)	 \cr 
GDS~J033213.53-274917.0	&  23.92	&  0.45	& - 	&-	&X	&faint	 \cr 
GDS~J033214.05-275124.5	&  22.69	&  0.74	&  1.220	&em.	&A	& [OII], MgII	 \cr 
GDS~J033214.33-274825.2	&  24.91	&-0.012	&  - 	&-	&X	& featureless continuum\cr  
GDS~J033214.38-274825.9	&  27.14	&  0.24	&  - 	&-	&X	& faint	 \cr 
GDS~J033214.69-275258.2	&  23.82	&  0.24	&  1.101	&em.	&A	& [OII]	 \cr 
GDS~J033214.71-275257.2	&  24.27	&  0.48	&  1.360	&em.	&B	& [OII]?	 \cr 
GDS~J033214.81-274600.0	&  24.24	&  0.54	& 1.370  	&em.	&C	& [OII]?	 \cr 
GDS~J033214.93-274659.8	&  23.75	&  0.12	& - 	&-	&X	&abs@7080A	 \cr 
GDS~J033215.01-274633.4	&  23.68	&  0.45	& 1.000 	&abs.	&C	& D4000 break?	 \cr 
GDS~J033215.09-275130.7	&  24.31	&  0.32	&  1.229	&em.	&A	& [OII]	 \cr 
GDS~J033215.88-274723.1	&  21.75	&  0.28	&  0.896	&em.	&A	& [OII], H$\beta$, TILT	 \cr 
GDS~J033216.02-274750.0	&  22.83	&  0.96	&  1.298	&comp.	&B	&CaHK, [OII](line6255?)	 \cr 
GDS~J033216.17-275241.4	&  22.55	&  0.78	&  1.094	&abs.	&B	&CaHK, g-band, MgI-noisy	 \cr 
GDS~J033216.28-274955.5	&  23.72	&  0.41	& - 	&-	&X	&featureless continuum \cr 
GDS~J033216.34-275013.4	&  23.25	&  0.52	&  1.046	&comp.	&A	& [OII](Sky-ABS), CaHK	 \cr 
GDS~J033216.37-275201.3	&  23.28	&  0.11	& - 	&-	&X	&bright, abs@6271,6982,7100	 \cr 
GDS~J033216.69-275239.0	&  22.57	&  0.88	&  1.045	&abs.	&A	&CaHK, g-band	 \cr 
GDS~J033216.91-274808.3	&  25.37	&  0.37	& - 	&-	&X	&faint	 \cr 
GDS~J033216.95-274519.3	&  23.61	&  0.78	&  1.303	&em.	&A	& [OII]	 \cr 
GDS~J033216.98-275102.4	&  23.54	&  0.40	&  0.991	&em.	&A	& [OII], [OIII], H$\beta$	 \cr 
GDS~J033217.29-274807.5	&  21.97	&  0.45	&  0.735	&abs.	&A	&CaHK	 \cr 
GDS~J033217.29-275113.2	&  22.73	&  0.85	& - 	&-	&X	&bad-row, featureless?	 \cr 
GDS~J033217.31-275025.0	&  23.50	&  0.19	&  1.612	&em.	&A	& [OII]	 \cr 
GDS~J033217.34-274844.3	&  24.88	&  0.35	&  1.107	&em.	&C	& [OII], faint	 \cr 
GDS~J033217.46-275234.8	&  21.93	&  0.96	&  1.100	&abs.	&A	&CaHK, MgI	 \cr 
GDS~J033217.47-274838.4	&  22.20	&  0.18	&  0.737	&em.	&A	& [OII], [OIII], H$\beta$	 \cr 
GDS~J033217.48-275248.0	&  21.82	&  1.08	&  1.095	&abs.	&A	&CaHK, g-band	 \cr 
GDS~J033217.56-274709.2	&  24.01	&  0.43	& - 	&-	&X	&faint	 \cr 
GDS~J033217.56-274810.1	&  24.36	&  0.13	&  0.542	&em.	&C	& [OII]?-faint	 \cr 
GDS~J033217.62-275228.5	&  21.18	&  0.78	&  1.098	&comp.	&A	&CaHK, MgI, g-band, AlII, [OII]	 \cr 
GDS~J033217.63-274811.8	&  22.57	&  0.52	&  0.735	&em.	&A	& [OII], CaHK, [OIII]	 \cr 
GDS~J033217.77-274603.0	&  24.23	&  0.91	& - 	&-	&X	&faint	 \cr 
GDS~J033217.78-274823.8	&  22.57	&  0.08	&  0.117	&em.	&B	&H$\alpha$,[OIII]	 \cr 
GDS~J033217.80-275256.9	&  24.30	&  0.50	&  1.044	&em.	&B	& [OII]?(SKY.ABS)	 \cr 
GDS~J033217.91-274122.7	&  22.10	&  0.96	&  1.041	&abs.	&A	&CaHK, MgI	 \cr 
GDS~J033217.94-274721.5	&  20.04	&  0.53	&  0.732	&abs.	&A	&CaHK, g-band, MgI, H$\delta$, H$\beta$, AlII	 \cr 
GDS~J033218.01-274718.5	&  19.41	&  0.61	&  0.735	&comp.	&A	& [OII], CaHK(noisy)	 \cr 
GDS~J033218.03-274850.3	&  23.20	&  0.18	&  0.297	&em.	&A	& [OIII], H$\beta$, H$\alpha$	 \cr 
GDS~J033218.07-274845.7	&  23.19	&  0.75	&  0.000	&star	&B	&star Flux-radius = 1.261 \cr 
GDS~J033218.19-274746.6	&  23.74	&  1.49	&  0.000	&star	&B	&star Flux-radius = 1.256	 \cr 
GDS~J033218.24-274744.0	&  23.63	&  0.39	& - 	&-	&X	&featureless continuum	 \cr 
GDS~J033218.58-274619.0	&  23.71	&  0.56	&  1.435	&em.	&A	& [OII]	 \cr 
GDS~J033218.61-274705.1	&  23.16	&  0.41	&  1.380	&em.	&A	& [OII], MgII	 \cr 
GDS~J033218.67-274915.7	&  24.18	&  0.28	& - 	&-	&X	&faint,line@8200A?	 \cr 
GDS~J033218.70-274919.8	&  22.69	&  0.77	&  1.038	&comp.	&A	& [OII], CaHK	 \cr 
GDS~J033218.78-274951.3	&  23.82	&  0.44	&  1.294	&em.	&A	& [OII], noisy \cr 
GDS~J033218.79-274820.8	&  23.41	&  0.30	&  0.999	&em.	&B	& [OII]	 \cr 
GDS~J033218.81-274908.5	&  23.90	&  0.43	&  1.128	&em.	&B	& [OII]?	 \cr 
GDS~J033218.81-274910.0	&  23.20	&  0.34	&  0.735	&comp.	&A	& [OII], CaHK	 \cr
GDS~J033219.15-274040.2	&  21.11	&  1.14	&  1.128	&abs.	&A	&CaHK, MgI, g-band, bright	 \cr 
GDS~J033219.23-274545.5	&  23.44	&  1.35	&  0.000	&star	&C	&star? Flux-radius = 1.239	 \cr 
GDS~J033219.30-275219.3	&  22.00	&  1.14	&  1.096	&abs.	&A	&CaHK, MgI, g-band	 \cr 
\hline
\label{tab:tblspec}
\end{tabular}
\end{table}
\setcounter{table}{1}
\begin{table}
\caption{Spectroscopic redshift catalog.}
\begin{tabular}{lcccccl}
\hline \hline
 ID(V1)       & $z_{850}$      & $(i_{775}-z_{850})$  & zspec         &class. & Quality & comments \\
\hline
GDS~J033219.43-274928.2	&  23.90	&  0.58	&  1.048	&em.	&B	& [OII](SKY.ABS)	 \cr 
GDS~J033219.48-274216.8	&  20.29	&  0.31	&  0.382	&comp.	&A	&CaHK, low-z	 \cr 
GDS~J033219.61-274831.0	&  21.87	&  0.10	&  0.671	&em.	&A	& [OII], [OIII], H$\beta$	 \cr 
GDS~J033219.68-275023.6	&  20.50	&  0.25	&  0.559	&comp.	&A	& [OII], CaHK, [OIII], H$\alpha$	 \cr 
GDS~J033219.77-274204.0	&  23.34	&  0.86	&  1.044	&comp.	&A	& [OII], CaHK	 \cr 
GDS~J033219.79-274609.9	&  24.11	&  0.88	&  1.221	&em.	&A	& [OII], MgI	 \cr 
GDS~J033219.79-274839.3	&  23.55	&  0.52	&  1.357	&em.	&A	& [OII]	 \cr 
GDS~J033219.89-274517.8	&  24.00	&  0.15	& - 	&-	&X	&bright, abs@8176,6716\AA	 \cr 
GDS~J033219.96-274449.8	&  22.35	&  0.24	&  0.783	&em.	&C	& [OII]?	 \cr 
GDS~J033219.97-274547.6	&  23.94	&  0.47	&  1.219	&em.	&A	& [OII]	 \cr 
GDS~J033219.99-274443.2	&  24.43	&  0.23	& - 	&-	&X	&faint	 \cr 
GDS~J033220.02-274104.2	&  21.51	&  0.52	&  0.682	&abs.	&A	&CaHK, g-band, MgI	 \cr 
GDS~J033220.11-275329.8	&  24.44	&  0.75	&  1.385	&em.	&C	& [OII]?	 \cr 
GDS~J033220.28-275233.0	&  22.22	&  0.92	&  1.119	&abs.	&A	&CaHK, MgI, g-band	 \cr 
GDS~J033220.29-274718.2	&  23.94	&  0.32	& - 	&-	&X	&faint, noisy	 \cr 
GDS~J033220.41-274641.7	&  24.15	&  0.35	&  1.227	&em.	&A	&[OII]	 \cr 
GDS~J033220.72-274932.6	&  24.16	&  0.80	& - 	&-	&X	&faint	 \cr 
GDS~J033220.91-275344.0	&  22.99	&  0.71	&  1.044	&em.	&B	& [OII](SKY.ABS), CaHK	 \cr 
GDS~J033221.22-274625.9	&  23.48	&  0.50	&  1.221	&em.	&A	& [OII]	 \cr 
GDS~J033221.57-274941.6	&  23.01	&  0.61	&  1.110	&em.	&A	& [OII]	 \cr 
GDS~J033221.63-274800.2	&  24.35	&  0.08	& - 	&-	&X	&featureless continuum	 \cr 
GDS~J033221.67-274056.0	&  22.75	&  0.65	&  1.045	&em.	&B	& [OII](SKY.ABS), CaHK	 \cr 
GDS~J033221.76-274442.1	&  20.86	&  0.18	&  0.295	&em.	&A	&H$\beta$, [OIII], H$\alpha$	 \cr 
GDS~J033221.81-274352.3	&  24.27	&  0.58	&  1.308	&em.	&B	& [OII]	 \cr 
GDS~J033221.84-274434.4	&  24.81	&  0.36	&  - 	        &-	&X	&faint	 \cr 
GDS~J033221.99-274655.9	&  20.42	&  0.47	&  0.670	&comp.	&A	& [OII], CaHK, g-band	 \cr 
GDS~J033222.18-274659.7	&  25.00	&  1.04	& - 	&-	&X	&faint	 \cr 
GDS~J033222.36-275018.4	&  22.82	&  0.27	&  0.736	&em.	&B	& [OII]	 \cr 
GDS~J033222.41-274858.0	&  24.19	&  0.46	&  1.383	&em.	&A	& [OII], MgI, MgII	 \cr 
GDS~J033222.47-275047.4	&  24.38	&  1.79	& 0.000		&star	&C	&star?  Flux-radius = 1.307	 \cr 
GDS~J033222.54-274603.8	&  24.32	&  0.25	& - 	&-	&X	&faint(line@9400\AA?)	 \cr 
GDS~J033222.58-274425.8	&  20.28	&  0.33	&  0.738	&comp.	&A	& [OII], H$\beta$, CaHK,MgI	 \cr 
GDS~J033222.93-274919.1	&  24.77	&  0.51	&  1.298	&em.	&B	& [OII]?	 \cr 
GDS~J033222.93-275104.6	&  22.89	&  0.38	&  0.905	&em.	&A	& [OII], H$\beta$	 \cr 
GDS~J033223.17-274219.6	&  23.82	&  0.78	& - 	&-	&X	&faint(abs@8157,9200)	 \cr 
GDS~J033223.18-274921.5	&  24.08	&  0.55	&  1.109	&em.	&B	& [OII]	 \cr 
GDS~J033223.26-275101.8	&  21.90	&  0.85	&  0.964	&abs.	&A	&CaHK, g-band	 \cr 
GDS~J033223.28-274744.7	&  23.94	&  0.24	&  0.764	&em.	&B	& [OII]	 \cr 
GDS~J033223.29-274742.6	&  24.31	&  0.41	&  1.092	&em.	&A	& [OII], CaK	 \cr 
GDS~J033223.40-274316.6	&  20.48	&  0.33	&  0.615	&comp.	&A	& [OII], CaHK, g-band, H$\beta$, [OIII]	 \cr 
GDS~J033223.45-274709.0	&  23.22	&  0.49	&  1.423	&em.	&A	& [OII], MgII	 \cr 
GDS~J033223.61-274601.0	&  22.57	&  1.06	&  1.033	&em.	&A	& [OII], red	 \cr 
GDS~J033223.61-275306.3	&  22.30	&  0.85	&  1.125	&abs.	&C	& CaHK? (noisy)	 \cr 
GDS~J033223.69-275324.4	&  20.41	&  0.40	&  0.532	&comp.	&A	& [OIII],  H$\alpha$	 \cr 
GDS~J033223.83-274639.4	&  24.19	&  0.84	&  1.222	&em.	&B	& [OII]	 \cr 
GDS~J033223.90-275326.2	&  23.81	&  1.07	&  - 	&-	&X	&faint	 \cr 
GDS~J033224.01-275039.0	&  22.74	&  0.98	&  1.094	&abs.	&A	&CaHK,MgI	 \cr 
GDS~J033224.08-275214.6	&  23.56	&  0.92	&  1.015	&em.	&C	& [OII]?, g-band?	 \cr 
GDS~J033224.11-274102.1	&  23.36	&  0.88	&  0.000	&star	&A	&star  Flux-radius = 1.247	 \cr 
GDS~J033224.20-274257.5	&  24.15	&  0.32	&  - 	&-	&X	&featureless continuum	 \cr 
GDS~J033224.20-274952.9	&  23.61	&  0.21	&  - 	&-	&X	&diffuse, faint	 \cr 
GDS~J033224.26-274126.4	&  20.18	&  0.44	&  0.533	&comp.	&A	& [OII], [OIII], H$\beta$, g-band	 \cr 
GDS~J033224.37-274315.2	&  24.63	&-0.06	&  1.271	&em.	&C	& [OII]?	 \cr 
GDS~J033224.39-274624.3	&  21.92	&  0.76	&  0.895	&abs.	&A	&CaHK, [OII] faint, (short-slit)	 \cr 
GDS~J033224.66-275051.9	&  22.40	&-0.28	&  0.272 	&em.	&C	&H$\alpha$, Mg?	 \cr 
GDS~J033224.72-274120.4	&  21.17	&  0.80	&  0.967	&abs.	&A	&CaHK, g-band, MgI, AlII	 \cr 
GDS~J033224.79-274912.9	&  24.90	&  1.61	&  0.000	&star	&C	&continuum+break,  Flux-radius = 1.284	 \cr 
GDS~J033224.85-275052.6	&  23.24	&  0.61	&  1.329	&em.	&A	& [OII]	 \cr 
GDS~J033224.90-274715.0	&  24.65	&  0.71	& - 	&-	&X	&faint	 \cr 
GDS~J033224.91-274923.7	&  23.87	&  0.08	& - 	&-	&X	&featureless bright continuum	 \cr 
\hline
\end{tabular}
\end{table}
\setcounter{table}{1}
\begin{table}
\caption{Spectroscopic redshift catalog.}
\begin{tabular}{lcccccl}
\hline \hline
 ID(V1)    & $z_{850}$      & $(i_{775}-z_{850})$  & zspec         &class. & Quality & comments \\
\hline
GDS~J033225.04-274718.2	&  23.86	&  0.51	&  1.357	&em.	&A	& [OII], MgII	 \cr 
GDS~J033225.10-274219.5	&  23.65	&  1.09	&  1.609	&em.	&B	& [OII]	 \cr 
GDS~J033225.19-274735.3	&  23.90	&  0.60	&  1.017	&em.	&B	& [OII], faint	 \cr 
GDS~J033225.20-275009.4	&  22.88	&  0.95	&  1.100	&abs.	&B	&CaHK, AlII	 \cr 
GDS~J033225.21-275335.0	&  21.17	&  0.72	&  0.833	&comp.	&A	& [OII], [OIII], CaHK, H$\delta$, (noisy)	 \cr 
GDS~J033225.35-274502.8	&  22.58	&  0.16	&  0.975	&em.	&A	& [OII], H$\beta$, [OIII]	 \cr 
GDS~J033225.47-274327.6	&  20.14	&  0.51	&  0.668	&abs.	&A	&CaHK, [OII], AlII	 \cr 
GDS~J033225.48-275211.6	&  23.68	&  0.39	&  1.312	&em.	&B	& [OII]	 \cr 
GDS~J033225.54-275209.1	&  22.93	&  0.45	&  0.955	&em.	&B	& [OII]?, noisy	 \cr 
GDS~J033225.55-275108.2	&  23.93	&  0.30	&  0.832	&em.	&A	& [OII], [OIII], H$\beta$	 \cr 
GDS~J033225.58-274529.0	&  24.33	&  0.15	&  0.667	&em.	&A	& [OII], [OIII]	 \cr 
GDS~J033225.69-274347.1	&  24.73	&  1.01	& - 	&-	&X	&faint	 \cr 
GDS~J033225.76-274347.0	&  23.02	&  1.11	& - 	&-	&X	&noisy, bad-row	 \cr 
GDS~J033225.77-274247.7	&  24.26	&  0.62	&  1.026	&em.	&B	&[OII](SKY.ABS)	 \cr 
GDS~J033225.79-274352.3	&  23.17	&  0.58	&  1.297	&em.	&A	& [OII], CaHK	 \cr 
GDS~J033225.86-275019.7	&  21.54	&  0.48	&  1.095	&em.	&A	&bright [OII], TILT	 \cr 
GDS~J033225.90-274341.2	&  18.92	&  0.88	&  0.000	&star	&C	& star?  Flux-radius = 1.296 \cr 
GDS~J033226.00-274150.6	&  21.75	&  0.23	&  0.545	&comp.	&A	& [OII], H$\beta$, CaHK	 \cr 
GDS~J033226.03-275147.7	&  22.98	&  0.53	&  1.242	&em.	&A	& [OII], MgI	 \cr 
GDS~J033226.16-274946.5	&  23.09	&  0.52	&  0.735	&comp.	&A	& [OII], CaHK	 \cr 
GDS~J033226.17-274603.6	&  24.25	&  0.54	&  1.219	&em.	&B	& [OII]	 \cr 
GDS~J033226.24-275005.6	&  23.68	&  0.73	&  1.096	&em.	&B	& [OII]  \cr 
GDS~J033226.26-274209.6	&  23.96	&  0.53	&  0.932	&em.	&B	& [OII]	 \cr 
GDS~J033226.31-274722.4	&  22.48	&  0.23	&  0.737	&em.	&A	& [OII], H$\beta$, [OIII]	 \cr 
GDS~J033226.32-274232.3	&  23.86	&  0.24	&  0.736	&em.	&B	& [OII]	 \cr 
GDS~J033226.40-274228.2	&  23.33	&  0.42	&  1.615	&em.	&A	& [OII], MgII	 \cr 
GDS~J033226.49-274035.5	&  19.60	&  0.19	& - 	&-	&X	&faint	 \cr 
GDS~J033226.64-274028.2	&  21.22	&  0.21	&  0.310	&em.	&A	&low-z, H$\alpha$, [OIII]	 \cr 
GDS~J033226.66-274025.1	&  21.70	&  0.29	&  1.042	&em.	&C	& [OII]?(SKY.ABS)	 \cr 
GDS~J033226.66-274029.8	&  22.06	&  0.89	&  1.040	&abs.	&A	&CaHK, g-band, MgI	 \cr 
GDS~J033226.67-274758.8	&  21.81	&  0.20	&  0.628	&em.	&C	&H$\beta$, [OIII]?	 \cr 
GDS~J033226.67-274834.8	&  23.59	&  0.55	&  0.905	&abs.	&C	&D4000break?	 \cr 
GDS~J033226.84-274545.3	&  23.47	&  0.52	&  1.306	&em.	&A	& [OII]	 \cr 
GDS~J033226.89-274541.9	&  23.72	&  0.13	&  0.338	&em.	&A	& [OII], H$\beta$, [OIII], H$\alpha$	 \cr 
GDS~J033226.92-274239.8	&  22.95	&  0.37	& - 	&-	&X	& featureless?	 \cr 
GDS~J033227.02-274407.2	&  22.15	&  0.84	&  1.128	&comp.	&A	& [OII], CaHK	 \cr 
GDS~J033227.05-275318.4	&  22.25	&  0.70	&  1.103	&comp.	&A	& [OII], CaHK	 \cr 
GDS~J033227.07-274404.7	&  22.24	&  0.16	&  0.739	&em.	&A	& [OII], H$\beta$, [OIII]	 \cr 
GDS~J033227.11-274922.0	&  22.88	&-0.01	&  0.559	&em.	&A	& [OII], H$\beta$, [OIII], H$\alpha$	 \cr 
GDS~J033227.17-274957.8	&  23.26	&  0.51	&  1.293	&em.	&A	& [OII], CaHK	 \cr 
GDS~J033227.36-274204.8	&  21.26	&  0.53	&  0.735	&comp.	&A	&CaHK, MgI, AlII, g-band	 \cr 
GDS~J033227.58-274051.7	&  23.43	&  0.94	& 1.070 	&em.	&C	&faint, [OII]?	 \cr 
GDS~J033227.70-274043.7	&  21.64	&  0.86	&  0.968	&abs.	&A	&CaHK, g-band, AlII	 \cr 
GDS~J033227.72-275040.8	&  21.42	&  0.74	&  1.097	&comp.	&A	& [OII], CaHK	 \cr 
GDS~J033227.84-274136.8	&  21.97	&  0.91	&  1.043	&abs.	&A	&CaHK, g-band, MgI, AlII	 \cr 
GDS~J033227.88-275140.4	&  20.26	&  0.40	&  0.521	&abs.	&A	&CaHK, g-band	 \cr 
GDS~J033228.09-275202.4	&  20.29	&  0.41	&  0.560	&abs.	&A	&g-band, Na, Mg, [OIII]	 \cr 
GDS~J033228.42-274700.2	&  24.31	&  0.19	& - 	&-	&X	&faint, abs@7060	 \cr 
GDS~J033228.44-274703.7	&  20.86	&  0.47	& - 	&-	&X	&noisy	 \cr 
GDS~J033228.45-274419.3	&  22.78	&  0.50	&  1.135	&em.	&A	& [OII], MgII	 \cr 
GDS~J033228.48-274059.6	&  23.82	&  0.49	& - 	&-	&X	&featureless continuum	 \cr 
GDS~J033228.56-274055.7	&  25.44	&  0.07	& - 	&-	&X	&faint	 \cr 
GDS~J033228.84-274132.7	&  25.43	&  0.01	&  4.800	&em.	&C	& $Ly_{\alpha}$? No continuum \cr 
GDS~J033228.88-274129.3	&  20.72	&  0.53	&  0.733	&comp.	&A	& [OII], CaHK, MgI, g-band, H$\delta$, H$\beta$ \cr 
GDS~J033228.94-274600.6	&  23.82	&  0.98	& - 	&-	&X	&abs@8555,8150,9200?	 \cr 
GDS~J033228.94-274128.2$\dag$    &  -    	&  -	&  4.882        &em.	&B	&$Ly_{\alpha}$, (SiIV?) \cr
GDS~J033228.99-274908.4	&  20.56	&  0.73	&  1.095	&abs.	&A	&CaHK, MgI, g-band	 \cr 
GDS~J033229.07-274153.1	&  24.27	&  0.44	& - 	&-	&X	&faint	 \cr 
GDS~J033229.22-274707.6	&  20.65	&  0.51	&  0.668	&abs.	&A	&CaHK, g-band	 \cr 
GDS~J033229.32-274054.0	&  23.74	&  0.74	& - 	&-	&X	&short-slit	 \cr 
\hline
\multicolumn{7}{l}
{$\dag$ not present in the catalog v1.0}\\
\end{tabular}
\end{table}
\setcounter{table}{1}
\begin{table}
\caption{Spectroscopic redshift catalog.}
\begin{tabular}{lcccccl}
\hline \hline
  ID(V1)     & $z_{850}$      & $(i_{775}-z_{850})$  & zspec         &class. & Quality & comments \\
\hline
GDS~J033229.35-275048.5	&  20.10	&  0.30	&  0.415	&abs.	&A	&H$\beta$, Mg, Na	 \cr 
GDS~J033229.48-274036.7	&  24.04	&  0.42	&  1.221	&em.	&C	& [OII]?	 \cr 
GDS~J033229.63-274511.3	&  24.07	&  0.70	&  1.033	&em.	&A	& [OII], MgI, CaH	 \cr 
GDS~J033229.65-274524.7	&  24.20	&  0.73	& - 	&-	&X	&faint	 \cr 
GDS~J033229.71-274507.2	&  22.24	&  0.17	&  0.736	&em.	&A	& [OII], H$\beta$, [OIII]	 \cr  
GDS~J033229.75-275147.1	&  23.25	&  0.54	&  1.315	&em.	&A	& [OII]	 \cr 
GDS~J033229.85-274520.5	&  21.01	&  0.56	&  0.953	&em.	&A	& [OII], AlII	 \cr 
GDS~J033229.87-274317.7	&  22.73	&  0.79	&  1.097	&comp.	&B	&CaHK, faint [OII]	 \cr 
GDS~J033229.99-274322.6	&  24.42	&  0.98	& - 	&-	&X	&faint	 \cr 
GDS~J033230.03-275026.8	&  23.43	&  0.83	&  1.005	&abs.	&B	&CaHK, MgII, g-band	 \cr 
GDS~J033230.06-274523.5	&  21.81	&  0.50	&  0.955	&em.	&A	& [OII]	 \cr 
GDS~J033230.07-274319.0	&  21.59	&  0.64	&  1.101	&comp.	&B	& [OII], CaHK	 \cr 
GDS~J033230.09-275100.3	&  20.80	&  0.52	&  0.733	&abs.	&A	&CaHK, g-band, MgI, AlII, H$\delta$, H$\beta$	 \cr 
GDS~J033230.23-274519.9	&  23.14	&  0.09	&  0.523	&em.	&A	& [OII]	 \cr 
GDS~J033230.34-274523.6	&  21.92	&  0.07	&  1.223	&comp.	&A	&faint [OII], CaHK, MgI	 \cr 
GDS~J033230.51-275004.4	&  23.85	&  1.03	& - 	&-	&X	&faint	 \cr 
GDS~J033230.70-274928.7	&  22.69	&  0.16	& - 	&-	&X	&faint	 \cr 
GDS~J033230.71-274617.2	&  22.24	&  0.66	&  1.307	&em.	&A	& [OII], TILT \cr 
GDS~J033230.75-274306.9	&  23.25	&  0.25	&  0.860	&em.	&A	& [OII], H$\beta$	 \cr 
GDS~J033230.83-274931.8	&  23.19	&  0.79	& - 	&-	&X	&abs@7150,9056? (noisy)	 \cr 
GDS~J033230.85-274621.7	&  21.60	&  0.74	&  1.018	&comp.	&A	&CaHK, [OII]	 \cr 
GDS~J033230.98-274434.9	&  23.71	&  0.50	&  1.222	&em.	&A	& [OII]	 \cr 
GDS~J033231.04-274050.2	&  21.74	&  0.77	&  1.037	&abs.	&A	&CaHK, g-band, MgI	 \cr 
GDS~J033231.22-274052.2	&  22.86	&  0.58	&  1.333	&em.	&B	& [OII], noisy	 \cr 
GDS~J033231.22-274532.7	&  22.98	&  1.00	&  1.097	&abs.	&A	&CaHK, g-band	 \cr 
GDS~J033231.28-274820.2	&  23.10	&  0.65	&  1.173	&comp.	&A	& [OII]	 \cr 
GDS~J033231.42-274324.1	&  23.01	&  0.82	&  1.025	&em.	&B	& [OII](SKY.ABS)	 \cr 
GDS~J033231.45-274435.0	&  17.90	&  0.49	&  0.000	&star	&A	&star  Flux-radius = 1.299	 \cr 
GDS~J033231.55-275028.8	&  23.96	&  0.68	& - 	&-	&X	&faint red	 \cr 
GDS~J033231.65-274504.8	&  23.16	&  0.54	&  1.098	&em.	&A	& [OII], MgI, CaHK,  H$\delta$,  H$\gamma$	 \cr 
GDS~J033232.04-274451.7	&  21.59	&  0.68	&  0.895	&abs.	&A	&CaHK	 \cr 
GDS~J033232.08-274119.4	&  22.59	&  0.53	&  1.036	&em.	&B	& [OII](SKY.ABS)	 \cr 
GDS~J033232.08-274155.2	&  23.00	&  0.39	&  0.960	&abs.	&C	&abs@7000,6400,7800	 \cr 
GDS~J033232.12-274359.3	&  24.66	&  0.18	& - 	&-	&X	&faint (line at 6217A?)  \cr 
GDS~J033232.14-274349.9	&  23.12	&  0.36	&  0.973	&em.	&A	& [OII]	 \cr 
GDS~J033232.32-274343.6	&  22.80	&  0.07	&  0.533	&em.	&C	& [OIII]	 \cr 
GDS~J033232.33-274345.8	&  23.95	&  0.42	&  1.025	&em.	&B	& [OII]	 \cr 
GDS~J033232.58-275053.9	&  21.61	&  0.21	&  0.669	&em.	&B	& [OII], H$\beta$	 \cr 
GDS~J033232.73-274538.8	&  21.69	&  0.31	& - 	&-	&X	&short-slit(line5900A)	 \cr 
GDS~J033232.73-275102.5	&  20.78	&  0.58	&  0.735	&abs.	&A	& CaHK, g-band	 \cr 
GDS~J033232.94-274543.9$\dag$	&  -	        &-	&  0.957	&em.	&A	& [OII]	 \cr 
GDS~J033232.96-274545.7	&  19.84	&  0.33	&  0.366	&abs.	&A	& N[II]+abs.spec. \cr 
GDS~J033233.00-275030.2	&  20.85	&  0.44	&  0.669	&em.	&A	& [OII], [OIII], H$\beta$	 \cr 
GDS~J033233.01-274829.4	&  22.80	&  0.12	&  0.664	&em.	&A	& [OIII], H$\beta$	 \cr 
GDS~J033233.02-274547.4	&  21.13	&  0.63	&  0.953	&abs.	&B	& CaHK, [OII]	 \cr 
GDS~J033233.08-275123.9	&  21.39	&  0.47	&  0.735	&abs.	&A	& CaHK	 \cr 
GDS~J033233.25-274117.4	&  24.06	&  0.86	& - 	&-	&X	&continuum, faint	 \cr 
GDS~J033233.28-274236.0	&  24.55	&  1.09	&  1.215 	&em.	&B	& [OII], faint red (XCDFS265)\cr 
GDS~J033233.41-274230.5	&  23.37	&  0.06	&  0.975	&em.	&B	& [OII]	 \cr 
GDS~J033233.71-274210.2	&  23.71	&  0.46	&  1.043	&em.	&B	& [OII](SKY.ABS), TILT	 \cr 
GDS~J033233.82-274410.0	&  21.11	&  0.42	&  0.667	&abs.	&A	& CaHK, g-band	 \cr 
GDS~J033233.85-274600.2	&  23.82	&  1.05	&  1.910	&abs.	&B	& MgII	 \cr 
GDS~J033234.00-274412.1	&  23.70	&  0.53	&  0.896	&em.	&B	& [OII], CaHK?	 \cr 
GDS~J033234.05-274937.8	&  22.91	&  0.71	&  0.832	&comp.	&A	& [OII], CaHK	 \cr 
GDS~J033234.08-274222.3	&  23.91	&  0.72	&  1.476	&em.	&B	& [OII]	 \cr 
GDS~J033234.82-274835.5	&  22.94	&  0.66	&  1.245	&em.	&A	& [OII], CaHK	 \cr 
GDS~J033234.85-274640.4	&  22.70	&  0.61	&  1.099	&em.	&A	& [OII]	 \cr 
GDS~J033235.08-274615.7	&  23.11	&  0.63	&  1.316	&em.	&A	& [OII]	 \cr 
GDS~J033235.11-275009.0	&  24.17	&  0.44	&  1.295	&em.	&A	& [OII]	 \cr 
\hline
\multicolumn{7}{l}
{$\dag$ not present in the catalog v1.0}\\
\end{tabular}
\end{table}
\setcounter{table}{1}
\begin{table}
\caption{Spectroscopic redshift catalog.}
\begin{tabular}{lcccccl}
\hline \hline
  ID(V1)     & $z_{850}$      & $(i_{775}-z_{850})$  & zspec         &class. & Quality & comments \\
\hline
GDS~J033235.19-275103.4	&  24.41	&  0.23	&  0.981	&em.	&B	& [OII]?	 \cr 
GDS~J033235.26-275104.8	&  22.79	&  0.44	&  0.734	&abs.	&A	&CaHK, H$\delta$, MgI	 \cr 
GDS~J033235.78-274627.5	&  22.76	&  0.89	&  1.094	&comp.	&A	& [OII], CaHK	 \cr 
GDS~J033235.79-274734.7	&  23.65	&  0.88	&  1.223	&em.	&A	& [OII], [NeIII]	 \cr 
GDS~J033236.04-275004.3	&  23.40	&  0.81	&  1.612	&em.	&A	& [OII], MgII	 \cr 
GDS~J033236.39-274747.0	&  23.18	&  0.17	& - 	&-	&X	&abs@6500,6586	 \cr 
GDS~J033236.43-274750.6	&  22.41	&  0.13	&  0.127	&em.	&B	& H$\alpha$, [OI]6300\AA, Na \cr 
GDS~J033237.19-274608.1	&  20.90	&  1.07	&  1.096	&abs.	&A	&CaHK	 \cr 
GDS~J033237.26-274610.3	&  22.17	&  0.19	&  0.736	&comp.	&A	& [OII], CaHK	 \cr 
GDS~J033237.56-274646.7	&  24.89	&  0.78	& - 	&-	&X	&abs@8195, em@8949?	 \cr 
GDS~J033238.27-274604.0	&  24.46	&  1.07	& - 	&-	&X	&faint	 \cr 
GDS~J033238.49-274702.4	&  21.26	&  0.76	&  0.953	&abs.	&A	&CaHK, MgI	 \cr 
GDS~J033239.01-274722.7	&  24.40	&  0.40	& - 	&-	&X	&faint	 \cr 
GDS~J033239.35-275016.3	&  22.05	&  0.42	& - 	&-	&X	&smoothly-red	 \cr 
GDS~J033239.56-274851.7	&  22.55	&  0.81	&  0.000	&star	&C	&star  Flux-radius = 1.269	 \cr 
GDS~J033239.60-274909.6	&  20.72	&  0.89	&  0.980	&abs.	&A	&CaHK, g-band, AlII	 \cr 
GDS~J033239.64-274709.1	&  22.70	&  0.99	&  1.317	&comp.	&A	& [OII], CaHK, MgI	 \cr 
GDS~J033239.67-274850.6	&  24.55	&  0.21	& - 	&-	&X	&featureless continuum	 \cr 
GDS~J033239.99-275114.2	&  23.75	&  0.62	& - 	&-	&X	&lines:8210,8800?	 \cr 
GDS~J033240.01-274815.0	&  25.33	&  1.44	&  5.828	&em.	&A	&$Ly_{\alpha}$	 \cr 
GDS~J033240.67-275032.3	&  24.56	&  0.21	& - 	&-	&X	&bad-row, faint	 \cr 
GDS~J033240.79-275035.1	&  21.69	&  0.12	&  0.213	&em.	&A	&H$\alpha$, S[II], (2d-order-light)	 \cr 
GDS~J033240.92-274823.8	&  24.00	&  0.45	&  1.244	&em.	&B	& [OII]?	 \cr 
GDS~J033241.21-274932.4	&  24.26	&  0.94	& - 	&-	&X	&faint	 \cr 
GDS~J033241.41-274457.5	&  23.02	&  0.80	& - 	&em.	&X	&em.lines@6815,8208	 \cr 
GDS~J033241.48-274440.4	&  23.00	&  0.84	&  1.296	&em.	&B	& [OII], faint	 \cr 
GDS~J033241.59-275003.0	&  22.32	&  0.66	&  0.000	&star	&B	&compact,  Flux-radius = 1.279	 \cr 
GDS~J033241.67-274448.7	&  23.83	&  0.14	& - 	&-	&X	&faint, short-slit	 \cr 
GDS~J033241.76-274619.4	&  20.93	&  0.35	&  0.333	&em.	&C	&low-z, H$\alpha$	 \cr 
GDS~J033242.07-274911.6	&  23.42	&  1.30	&  0.000	&star	&B	&star  Flux-radius = 1.253	 \cr 
GDS~J033242.21-274953.9	&  23.83	&  0.37	&  1.377	&em.	&A	& [OII], MgI	 \cr 
GDS~J033242.25-274625.4	&  22.50	&  0.63	&  1.288	&em.	&A	& [OII], MgII	 \cr 
GDS~J033242.32-274950.3	&  20.33	&  0.29	& - 	&-	&X	&noisy	 \cr 
GDS~J033242.38-274707.6	&  23.18	&  1.11	&  1.314	&abs.	&B	& [OII], CaHK	 \cr 
GDS~J033242.56-274550.2	&  22.21	&  0.16	&  0.218	&em.	&A	& [OIII], H$\beta$, H$\alpha$	 \cr 
GDS~J033242.97-274649.9	&  23.44	&  0.89	& - 	&-	&X	&faint	 \cr 
GDS~J033244.18-274729.4	&  23.55	&  0.44	&  1.220	&em.	&A	& [OII]	 \cr 
GDS~J033244.20-274733.5	&  21.53	&  0.20	&  0.737	&em.	&A	& [OII], [OIII], H$\beta$	 \cr 
GDS~J033244.23-275039.5	&  24.72	&  0.40	&  1.122	&em.	&B	& [OII]?	 \cr 
GDS~J033244.29-275009.7	&  22.62	&  0.86	&  1.038	&abs.	&A	&CaHK, g-band	 \cr 
GDS~J033244.43-274641.8	&  22.55	&  0.13	&  0.215	&em.	&C	&H$\alpha$?	 \cr 
GDS~J033244.62-274632.2	&  23.20	&  0.44	&  1.426	&em.	&A	& [OII], MgII	 \cr 
GDS~J033244.80-274920.6	&  23.69	&  0.84	& - 	&-	&X	&faint	 \cr 
GDS~J033245.15-274940.0	&  21.92	&  1.04	&  1.123	&abs.	&A	&CaHK, g-band, MgI	 \cr 
GDS~J033245.21-274858.0	&  24.65	&  0.51	&  1.463	&em.	&A	& [OII]	 \cr 
GDS~J033245.90-274517.2	&  23.79	&  0.39	&  1.036	&em.	&A	& [OII], CaHK	 \cr 
GDS~J033247.45-274603.9	&  23.98	&  0.51	& - 	&-	&X	&faint	 \cr 
GDS~J033248.56-274504.6	&  22.19	&  0.88	&  1.115	&abs.	&C	&CaHK, noisy	 \cr 
GDS~J033249.04-275015.5	&  22.36	&  0.80	&  1.122	&em.	&B	& [OII]	 \cr 
GDS~J033249.09-274519.2	&  21.46	&  0.37	& - 	&-	&X	& featureless continuum	 \cr 
GDS~J033249.11-274524.2	&  21.81	&  0.60	&  1.094	&em.	&A	& [OII], H$\beta$-faint	 \cr 
GDS~J033249.49-274534.2	&  23.94	&  0.17	&  1.609	&em.	&A	& [OII]	 \cr 
GDS~J033249.85-274757.8	&  22.78	&  0.67	&  1.146	&em.	&B	& [OII]?	 \cr 
GDS~J033250.69-274732.2	&  23.51	&  -0.02	& - 	&-	&X	&em@7100A, abs8100A	 \cr 
GDS~J033251.34-274742.7	&  24.09	&  0.76	&  1.298	&em.	&B	& [OII], noisy	 \cr 
GDS~J033251.57-275044.7	&  22.48	&  0.52	&  0.980	&comp.	&A	& [OII], H$\beta$	 \cr 
GDS~J033252.87-275114.7	&  21.20	&  0.63	&  1.002	&comp.	&A	&CaHK, [OII]	 \cr 
GDS~J033252.88-275119.8	&  21.84	&  0.47	&  1.220	&em.	&A	& [OII], CaHK	 \cr 
GDS~J033253.01-275000.5	&  24.05	&  0.09	& - 	&-	&X	&faint	 \cr 
GDS~J033253.34-275104.6	&  25.50	&-0.08	&  0.912	&em.	&C	&noisy, [OII]?	 \cr 
GDS~J033255.00-275051.6	&  21.70	&  0.37	& - 	&-	&X	&featureless continuum	 \cr 
\hline
\end{tabular}
\end{table}

\section{Discussion}

\subsection{Reliability of the redshift - comparison with VVDS}

\begin{figure}
 \centering
 \includegraphics[width=\textwidth]{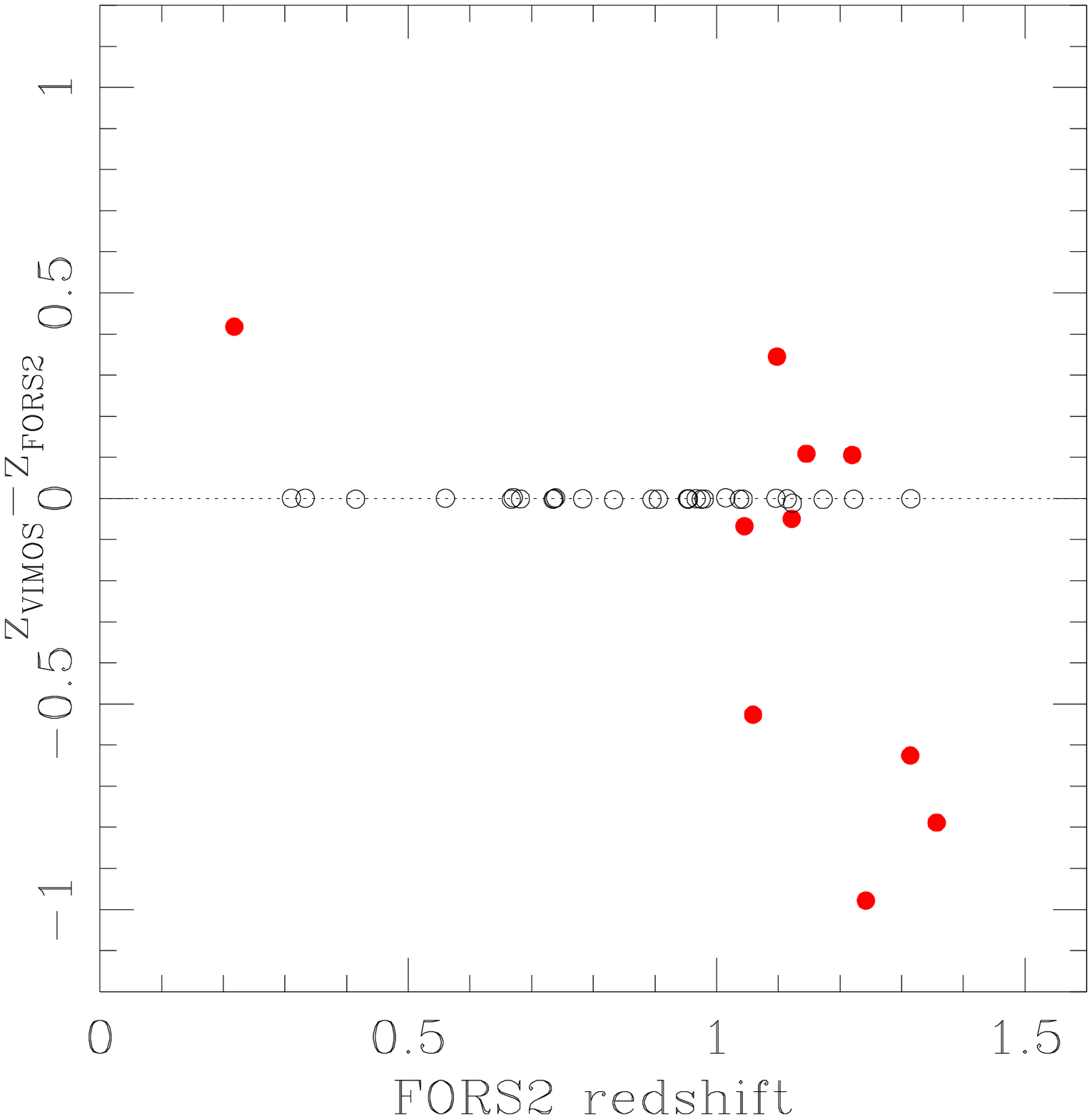}
 \caption{VVDS spectroscopic redshift versus FORS2 spectroscopic
 redshift. There are 39 galaxies in common between the VVDS 
 sample and the sample presented in
 this work. 10 cases show (filled symbols) discrepant redshift
 determination with $|dz|>$0.015.}
 \label{fig:VIMOS_vs_FORS2}
\end{figure}

A practical way to assess the reliability of the redshifts reported in
Table~\ref{tab:tblspec} is to compare the present results with 
independent measurements of other surveys. From this point of view the
recent release of the data of the VIMOS-VLT Deep Survey (VVDS, \cite{fevre04})
is particularly important. 
There are 39 VVDS objects in common with the first release of the 
FORS2 GOODS survey and Figure~\ref{fig:VIMOS_vs_FORS2} shows the comparison 
of the redshift determinations. The reliability level of the redshift 
measurements in the VVDS is indicated by a quality flag. Flags 2, 3, 4 
are the most secure with a confidence of 75$\%$, 95$\%$ and 100$\%$ respectively.
Flag 1 is an indicative measurement, flag 9 indicates that there is
only one secure emission line, and flag 0 indicates a measurement
failure with no features identified.

For 29 cases out of 39 (74$\%$) the agreement is very good, with a 
mean difference $<z_{FORS2}-z_{VVDS}>$ = 0.0016 $\pm$ 0.0021.

Assuming equipartition of the redshift uncertainties between FORS2 and
VVDS, we can estimate a $\sigma_{z} (FORS2) \simeq 0.0015$, in
reasonable agreement with the estimate of Sect.~4. In the following
we will assume a typical uncertainty of the redshift determinations of
the present survey  to be  $\sigma_{z} \simeq 0.001$ (excluding
``catastrophic'' discrepancies). 

Ten cases show ``catastrophic'' discrepancies, i.e. 
$<z_{FORS2}-z_{VVDS}>$ greater than 0.015 and are reported in 
Table~\ref{tab:FORS2vsVIMOS}. 

In the following we discuss case by case the origin of the discrepancy:
\begin{enumerate}
\item{GDS~J033214.05-275124.5:}
\begin{itemize} 
\item{FORS2: the emission line [O\,{\sc ii}]3727 and the absorption lines Ca H and K are detected in 
the FORS2 spectrum at z=1.220. The absorption line MgII 2798 is also present
at 6210\AA. The 4000\AA\ ~Balmer Break is also evident, quality flag ``A''.} 
\item{VVDS: the main emission feature in the VIMOS spectrum is identified with [O\,{\sc ii}]3727 
at z=1.325, quality flag 3. We note an absorption feature in the VIMOS spectrum
(without identification) at $\sim$6200\AA, consistent with the one measured in the 
FORS2 spectrum.}
\end{itemize}
\item{GDS~J033219.79-274839.3:}
\begin{itemize}
\item{FORS2: flat continuum with an evident emission line at $\sim$8784\AA. 
We interpret it as [O\,{\sc ii}]3727. No spectroscopic feature is observed 
at $\sim$ 6500\AA. Quality flag A.}
\item{VVDS: the main feature in the VIMOS spectrum is identified with 
[O\,{\sc iii}]5007 at z=0.568 (emission line at 7851\AA), quality flag 2.}
\end{itemize}
\item{GDS~J033221.67-274056.0:}
\begin{itemize}
\item{FORS2: the emission line [O\,{\sc ii}]3727 and the absorption lines Ca H and K are detected in the FORS2 
spectrum at $z=1.045$, quality flag ``B''. The [O\,{\sc ii}]3727 line is attenuated by the sky absorption band
at $\sim$7600\AA.}
\item{VVDS: Ca H and K are identified in the VIMOS spectrum at z=0.977, quality flag 1.}
\end{itemize}
\item{GDS~J033226.03-275147.7:}
\begin{itemize}
\item{FORS2: the emission line [O\,{\sc ii}]3727 (at 8356\AA) and the absorption lines Ca H and K, MgI
and B2630 are detected in the FORS2 spectrum at $z=1.242$, quality flag ``A''.}
\item{VVDS: the main feature is identified with H$\alpha$ at z=0.264 (at 8296\AA), quality flag 9.
No emission lines are present in the FORS2 spectrum at this wavelength.}
\end{itemize}
\item{GDS~J033231.65-274504.8:}
\begin{itemize}
\item{FORS2: the emission line [O\,{\sc ii}]3727 and the absorption lines Ca H and K, MgI, 
H$\delta$ and are detected in the FORS2 spectrum at $z=1.098$; the emission line 
H$\gamma$ is also detected at $\sim$9105\AA, quality flag ``A''.}
\item{VVDS: in the VIMOS spectrum a line is detected at $\sim$9105\AA, interpreted as 
[O\,{\sc ii}]3727 at z=1.443, quality flag 1.}
\end{itemize}
\item{GDS~J033232.32-274343.6:}
\begin{itemize}
\item{FORS2: for this object (at the border of the FORS2 field of
view) the spectrum starts at $\sim$6400\AA. We detect a weak emission line at 
$\sim$7654\AA\ (close to a sky absorption band), that we originally 
interpreted to be [O\,{\sc ii}]3727 at z$\sim$1.059, 
assigning to the redshift a quality flag ``C''.}
\item{VVDS: in the VIMOS spectrum an emission line
at $\sim$5713\AA\ is detected, interpreted as [O\,{\sc ii}]3727 at z=0.533 and quality flag 3
(the absorption feature Ca H at $\sim$6085\AA\ is also present). 
It is consistent with the interpretation [O\,{\sc ii}]3727 at $z=0.533$ with the FORS2
$\sim$ 7654\AA\ emission line identified as [O\,{\sc iii}]5007 at $z=0.533$. We have therefore
updated the entry in Table~\ref{tab:tblspec} to a redshift $z=0.533$.}
\end{itemize}
\item{GDS~J033242.38-274707.6:}
\begin{itemize}
\item{FORS2: this is a red object ($i_{775}-z_{850}$ = 1.11), we detect two clear 
absorption features in the $\sim$ 9100\AA\ sky free region interpreted as Ca H and K, 
faint [O\,{\sc ii}]3727 seems to be present, quality flag ``B''.}
\item{VVDS: red spectrum, Ca H and K are identified in the VIMOS spectrum 
at z=0.688, quality flag 2.}
\end{itemize}
\item{GDS~J033242.56-274550.2:}
\begin{itemize}
\item{FORS2: the emission lines [O\,{\sc iii}]5007 (at 6098\AA), H$\beta$ (at 5921 \AA) 
and H$\alpha$ (at 7994\AA) are detected in the FORS2 spectrum, z=0.218, quality flag ``A''.
}
\item{VVDS: the main emission feature (at 6094\AA) is identified with [O\,{\sc ii}]3727 
at z=0.635, quality flag 2.}
\end{itemize}
\item{GDS~J033249.04-275015.5:}
\begin{itemize}
\item{FORS2: the spectrum starts at $\sim$6400\AA. It shows continuum 
with a evident emission line at $\sim$7909\AA\ interpreted as [O\,{\sc ii}]3727 
(z=1.122), a discontinuity consistent with the 4000\AA\ ~Balmer Break is 
present, quality flag ``B''.}
\item{VVDS: the main feature in the VIMOS spectrum is an emission line at 7723\AA
 identified with [O\,{\sc ii}]3727 at z=1.072, quality flag 2.}
\end{itemize}
\item{GDS~J033249.85-274757.8:}
\begin{itemize}
\item{FORS2: object red with bright continuum, the emission line [O\,{\sc ii}]3727 
and the absorption lines MgII 2798 and $H\zeta$ are detected in the FORS2 spectrum, 
the 4000\AA\ ~Balmer Break is also evident, quality flag ``B''.}
\item{VVDS: flat continuum, the NeV absorption line is identified at z=1.254, 
quality flag 2.}
\end{itemize}
\end{enumerate}

\begin{table}
\centering \caption{Galaxies with discrepant redshifts between the
FORS2 and VVDS surveys.}
\begin{tabular}{lcccccc}
\hline \hline
 N & ID & z(FORS2) & QF(FORS2) & z(VVDS) & QF(VVDS) & z(FORS2)-z(VVDS)\\
\hline
 1 & GDS~J033214.05-275124.5 & 1.220 & A (em.) & 1.325 & 3 & -0.105 \cr
 2 & GDS~J033219.79-274839.3 & 1.357 & A (em.) & 0.568 & 2 & 0.789\cr 
 3 & GDS~J033221.67-274056.0 & 1.045 & B (em.) & 0.978 & 1 & 0.067\cr
 4 & GDS~J033226.03-275147.7 & 1.242 & A (em.) & 0.264 & 9 & 0.978\cr
 5 & GDS~J033231.65-274504.8 & 1.098 & A (em.) & 1.443 & 1 & -0.350 \cr
 6 & GDS~J033232.32-274343.6 & 1.059$\dag$ & C (em.) & 0.533 & 3 & 0.526 \cr
 7 & GDS~J033242.38-274707.6 & 1.314 & B (abs.) & 0.688 & 2 & 0.626\cr
 8 & GDS~J033242.56-274550.2 & 0.218 & A (em.) & 0.635 & 2 & -0.417 \cr
 9 & GDS~J033249.04-275015.5 & 1.122 & B (em.) & 1.072 & 2 & 0.05 \cr
 10 & GDS~J033249.85-274757.8 & 1.146 & B (em.) & 1.254 & 2 & -0.105 \cr 
\hline
\multicolumn{7}{l}
{$\dag$ Adopted FORS2 value has been updated to VVDS value in Table 2.}\\
\label{tab:FORS2vsVIMOS}
\end{tabular}
\end{table}
In summary, out of ten highly discrepant cases we have found only
one that can be ascribed to an evident error in the identification of the
features in the FORS2 spectrum (and the original quality flag for this
object was ``C''). We conclude that the probable fraction of
``catastrophic'' misidentifications in Table~\ref{tab:tblspec} is at most
a few percent.

\subsection{Reliability of the redshifts - diagnostic diagrams}

\begin{figure}
 \centering
 \includegraphics[width=\textwidth]{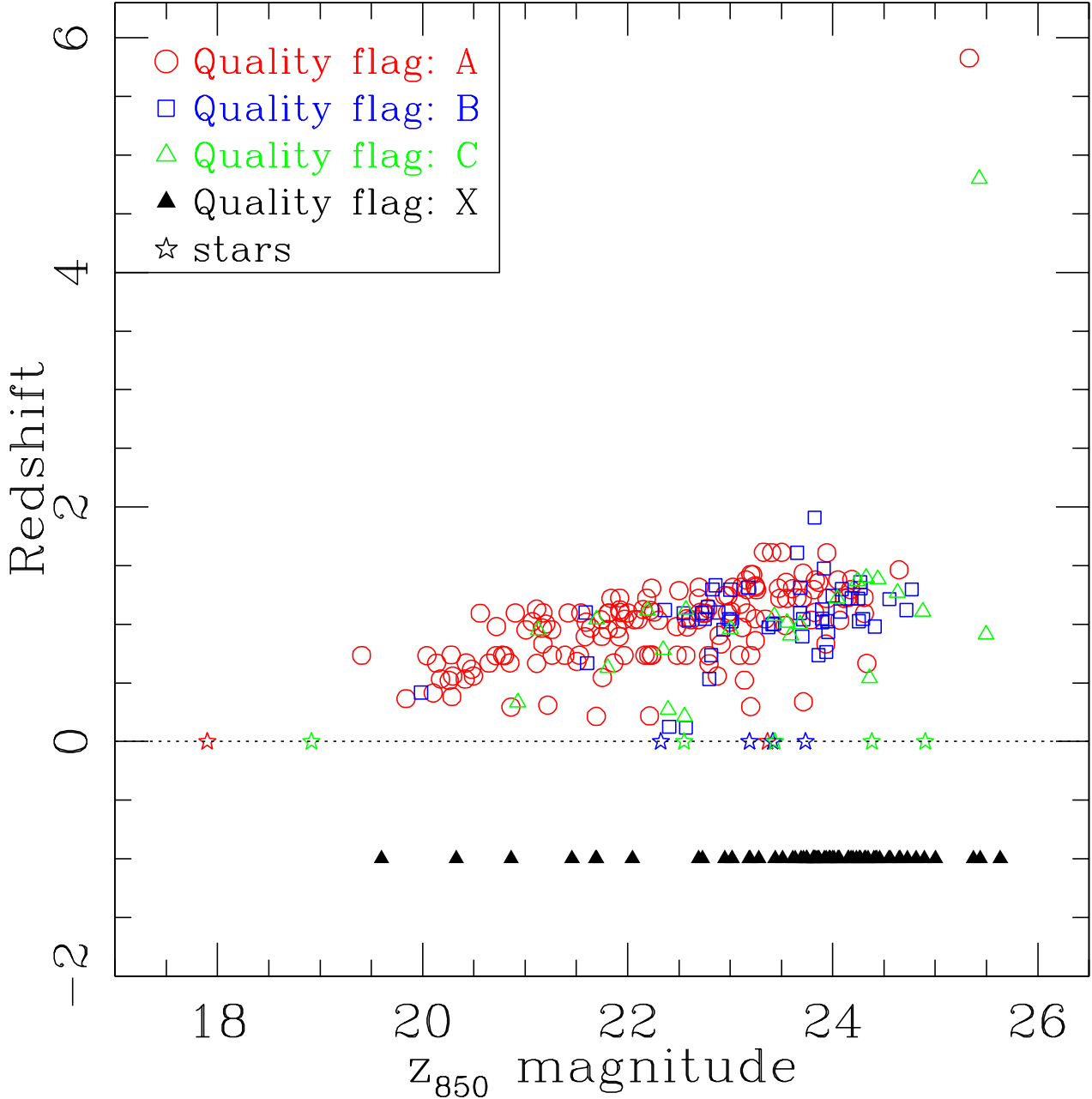}
 \caption{Spectroscopic redshift versus magnitude for the entire FORS2
sample (quality flag ``A'', ``B'', ``C'' and ``X''). Stars are
denoted by star-like symbols at zero redshift. Inconclusive spectra
are placed at $z=-1$.} 
\label{fig:z_vs_mag}
\end{figure}

\begin{figure}
 \centering
 \includegraphics[width=\textwidth]{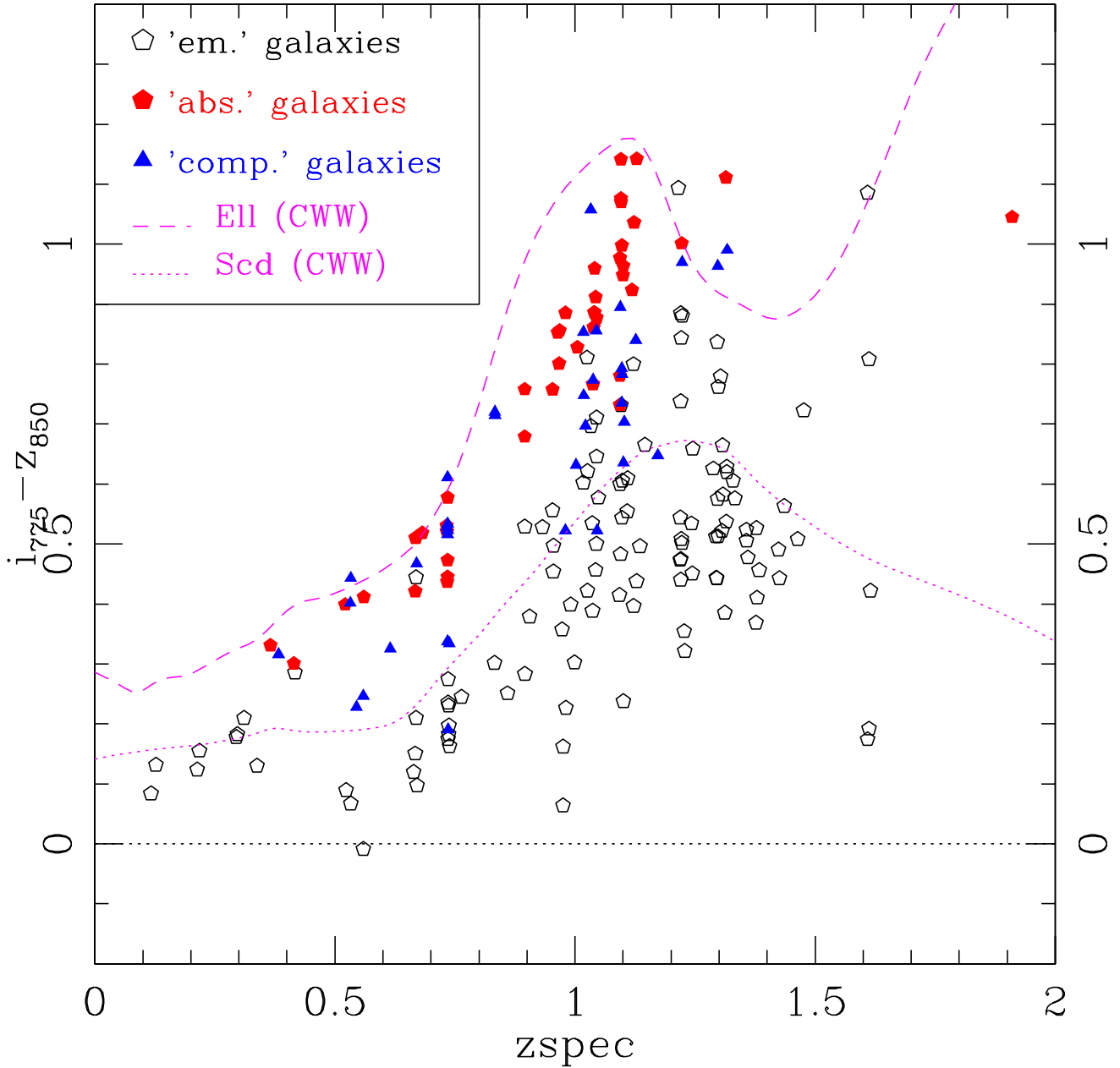}
 \caption{Color-redshift diagram of the spectroscopic sample.
Only redshifts with quality flag ``A'' and ``B'' have been selected. 
Filled pentagons symbols are objects identified with absorption
features only (``abs.'' sources), 
while open pentagons are objects showing only 
emission lines (``em.'' sources). 
The intermediate cases are shown by filled triangles (``comp.'' sources).
The long-dashed line and the short dashed line show the colors of a non-evolving 
$L^{\star}$ elliptical galaxy and an Scd galaxy, respectively, estimated 
integrating the spectral templates of \cite{cole80} through the ACS bandpasses.}
\label{fig:i_zVSzspec}
\end{figure}

\begin{figure}
 \centering
 \includegraphics[width=\textwidth]{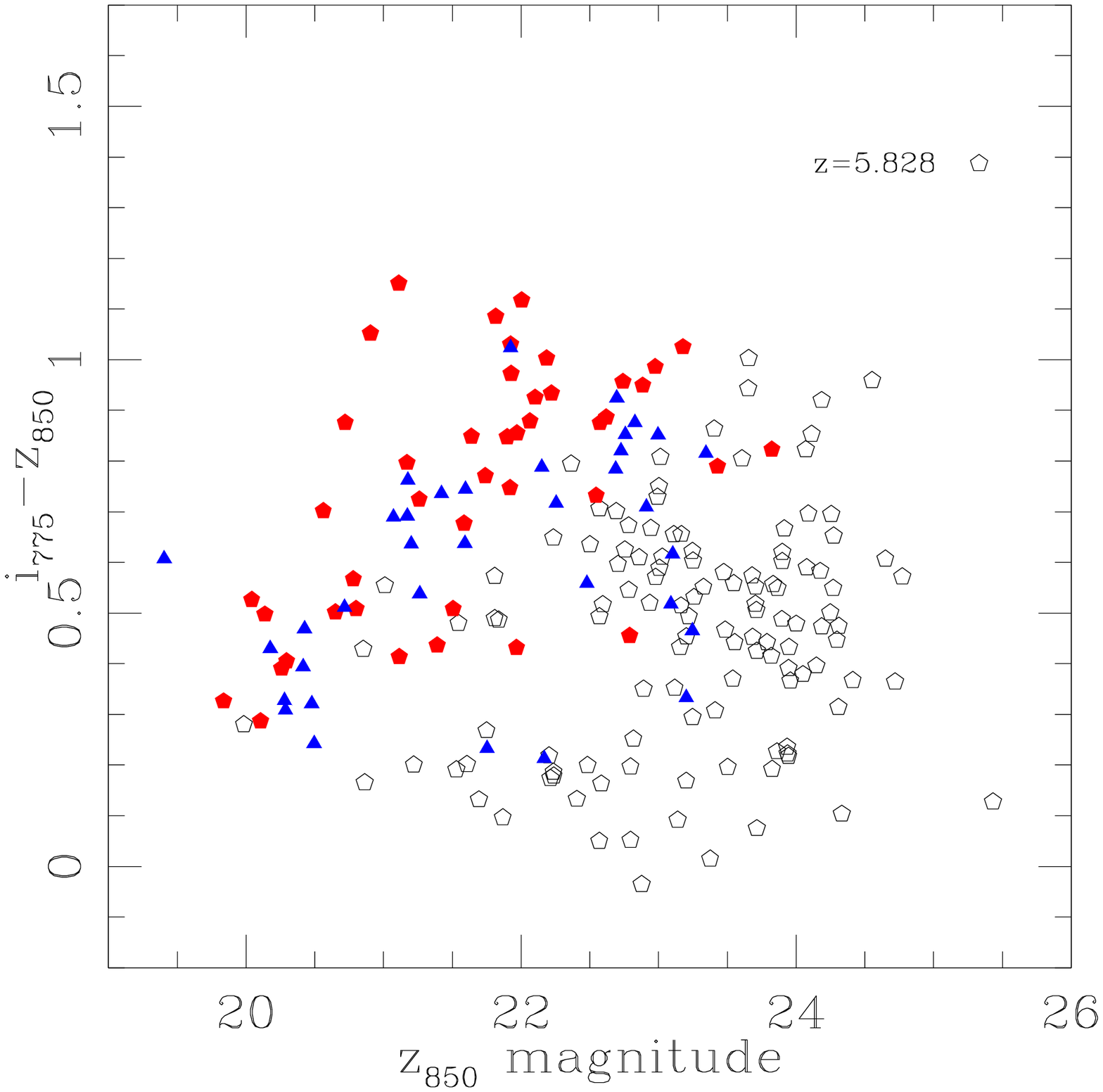}
 \caption{Color-magnitude diagram for the spectroscopic sample. 
Only redshifts with quality flag ``A'' and ``B'' have been selected. 
The symbols are the same as in Figure~\ref{fig:i_zVSzspec}.}
\label{fig:i_zVSmagz}
\end{figure}

As mentioned above, the photometric information and its relation with
the redshift provides useful indications about possible errors in the
redshift measurement and/or magnitude estimation.
The Figures~\ref{fig:z_vs_mag}, \ref{fig:i_zVSzspec} and
\ref{fig:i_zVSmagz} show the redshift-magnitude, the color-redshift
and the color-magnitude distributions for the spectroscopic sample
(the quality flag ``A'' and ``B'' have been selected in the
Figures~\ref{fig:i_zVSzspec} and \ref{fig:i_zVSmagz}, while all
sources have been plotted in the Figure~\ref{fig:z_vs_mag}). 
In figure~\ref{fig:i_zVSzspec} the two populations of
``emission-line''  and ``absorption-line'' (typically elliptical) galaxies 
are clearly separated.
The mean color of the ``absorption-line'' objects increases from $i_{775}-z_{850}$
= 0.46 $\pm$ 0.079 at $<z>$ = 0.6 to $i_{775}-z_{850}$ = 0.86 $\pm$ 0.18 at $<z>$
= 1.0, consistent but increasingly bluer than the colors of a non-evolving $L^{\star}$
elliptical galaxy (estimated integrating the spectral templates of \cite{cole80} 
through the ACS bandpasses).

The ``emission-line'' objects show in general a bluer $i_{775}-z_{850}$ color and
a broader distribution than the ``absorption-line'' sources: $i_{775}-z_{850}$ =
0.16 $\pm$ 0.13 at $<z>$ = 0.6 and $i_{775}-z_{850}$ = 0.52 $\pm$ 0.21 at $<z>$ =
1.1.
The broader distribution, with some of the ''emission-line'' objects
entering the  color regime of the ellipticals, 
is possibly explained by dust obscuration,
high metallicity or strong line emission in the $z_{850}$ band. 

\begin{table}
\centering \caption{Fractions of sources with different spectral features.}
\begin{tabular}{lccccc}
\hline \hline
 Spectral class& $z_{mean}$ & $z_{min}$ & $z_{max}$& Fraction \\
\hline
 emission  & 1.131 & 0.117 & 5.828 & 46$\%$ \cr
 absorption  & 0.950 & 0.366 & 1.910 & 16$\%$ \cr
 em. \& abs. & 0.897 & 0.382 & 1.317 & 12$\%$ \cr
 stars       & 0.000 & 0.000 & 0.000 & 4$\%$ \cr
 unclassified & -  & -  & -  & 22$\%$ \cr 
\hline
\label{tab:z_properties}
\end{tabular}
\end{table}
\subsection{Redshift distribution and Large Scale Structure}
\begin{figure}
 \centering
 \includegraphics[width=\textwidth]{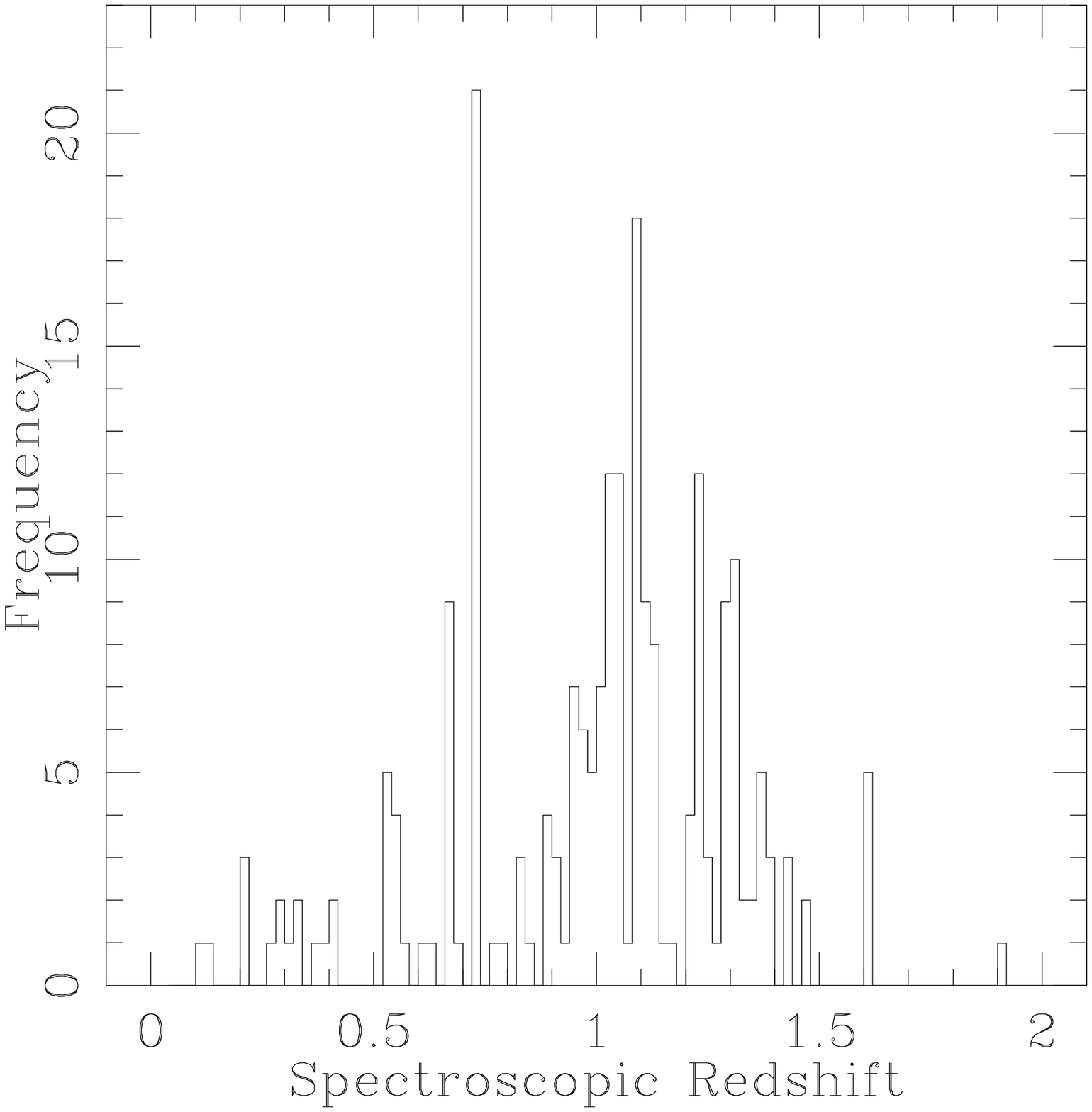}
 \caption{Redshift distribution for the spectroscopic sample with 
 quality A, B and C (23 redshift determinations out of 224 have quality C). 
Three objects at z$>$4 are not shown in the histogram.}
\label{fig:zdistr}
\end{figure}

Figure~\ref{fig:zdistr} shows the redshift distribution of the objects
observed in the present survey.  The majority of the sources are at
redshift around $\sim$1 (the median of the redshift distribution is at 1.04),
in agreement with the main criterion for the target selection (see
Sect.~2).  Table~\ref{tab:z_properties} shows the fraction of
determined redshifts as a function of the spectral features identified,
i.e. emission lines, absorption lines, emission \& absorption lines,
and no reliable spectral features (unclassified).  There are 49
galaxies identified with absorption lines only (mainly Ca H and K) in
the range of redshift between 0.4-1.3; an example is shown in
Figure~\ref{fig:reduction}. In 46$\%$ of the total sample we have
measured emission lines (mainly [O\,{\sc ii}]3727), many of them
entering the so-called ``spectroscopic desert'' up to z=1.61. 

\begin{figure}
 \centering
 \includegraphics[width=\textwidth]{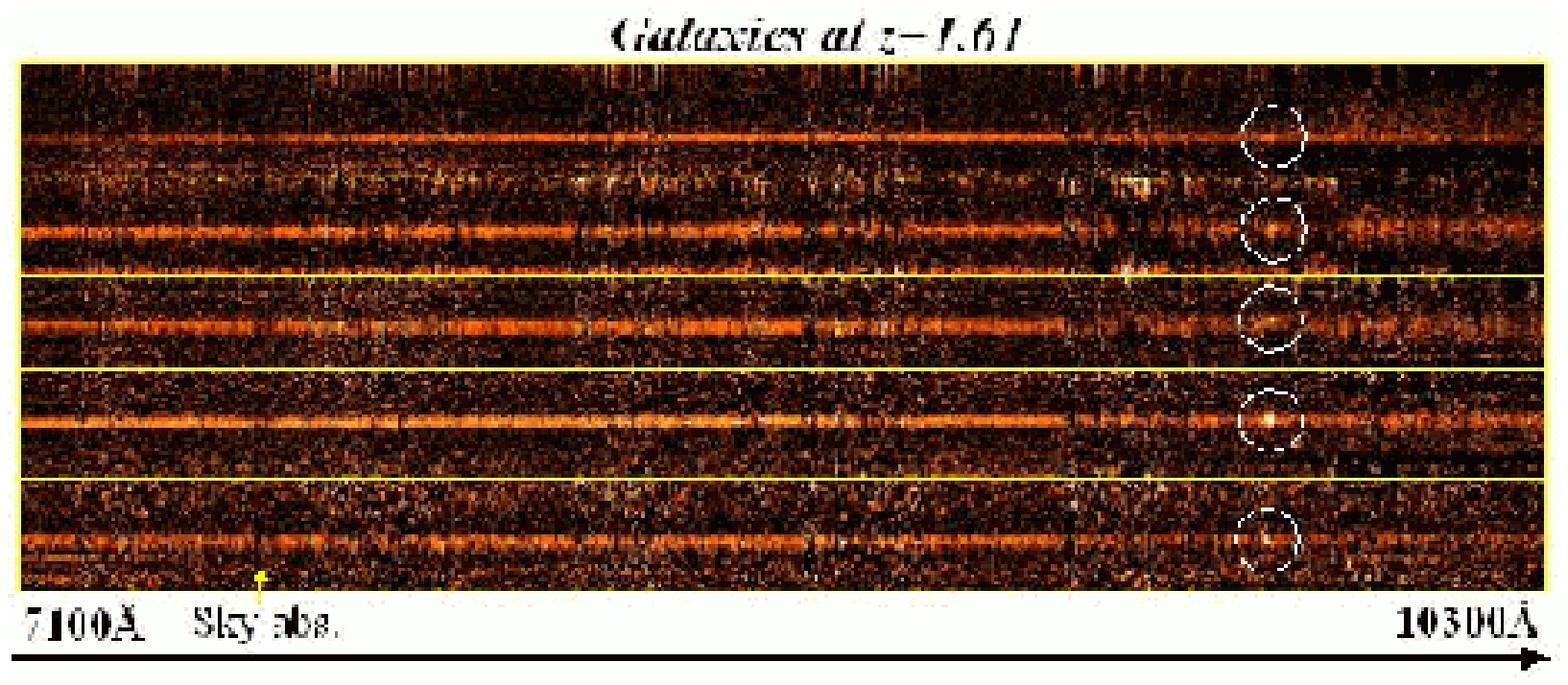}
 \caption{Two dimensional spectra of 5 galaxies at z=1.61. The
[O\,{\sc ii}]3727 emission line is marked with a circle at
9727.5\AA. The absorption sky feature ($\sim$7600\AA, A band) is
indicated with an arrow. It is worth to note the optimal red
sensitivity of FORS2.}
\label{fig:groupz1.61}
\end{figure}

The main peaks in the redshift distribution are at $z \sim 0.73$
(21 galaxies) and $1.1$ (25 galaxies).
Two concentrations at $z \sim 1.6$ (with 5 galaxies at the mean 
redshift $<z> = 1.612 \pm 0.003$, see the two dimensional spectra in 
Figure~\ref{fig:groupz1.61}) and $z \sim 0.67$ (9 galaxies) are also apparent.
The presence in the CDF-S of large scale structure, (LSS) at $z \sim 0.73$ 
and $z \sim 0.67$ is already known (\cite{cimatti02}, \cite{gilli03}, \cite{fevre04}).
The peak at $z \sim 1.1$ seems to be a new indication of large scale
structure, of the 25 galaxies in the range 1.09$<$z$<$1.11,
10 show emission lines, 9 are ellipticals and 6 are
intermediate-type galaxies. 

The significance of the LSS at $z=1.61$ is confirmed by:
\begin{enumerate}
\item the observations of 
Gilli et al. (2003) who found a peak in the redshift distribution of X-ray sources
at z=1.618 (5 galaxies) and measured a Poissonian probability 
of 3.8$\times 10^{-3}$ for a chance distribution ;
\item three more galaxies at $z = 1.605, 1.610, 1.615$ in the K20 survey
\cite{cimatti02}; 
\end{enumerate}
\begin{figure}
 \centering
 \includegraphics[width=13cm,height=15cm]{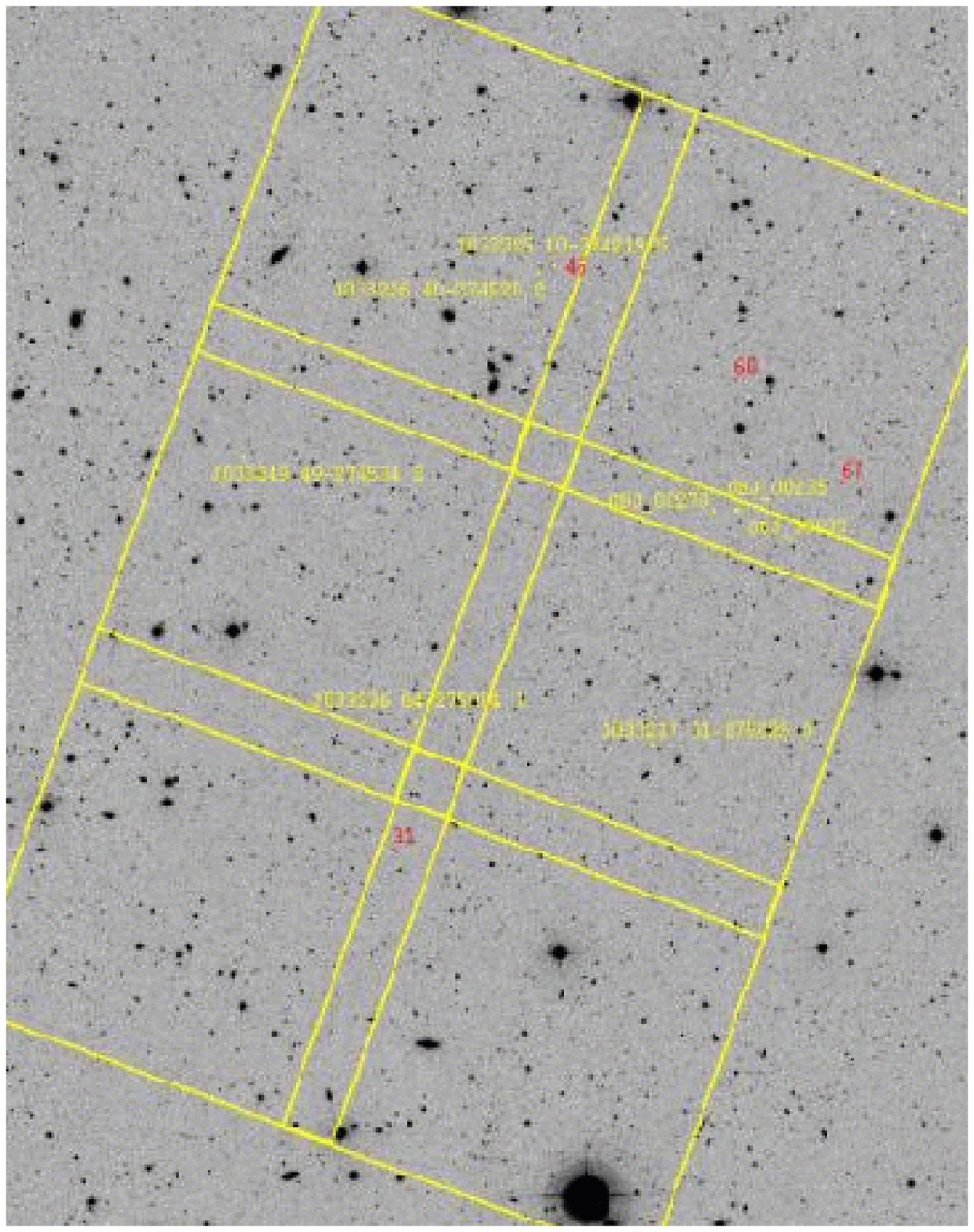} 
 \caption{The spatial distribution of the galaxies at $z \sim 1.61$ in
the CDF-S. The background image is an exposure in the $R$ band
obtained with the ESO wide-field imager (WFI). North is up and east on
the left. The squares represent
the FOV of the FORS2 pointings. The five FORS2 targets are given by their
coordinates, and the three K20 sources with the identifiers: 
OBJ$\_$00235, OBJ$\_$00237 and OBJ$\_$00270. The numbers 31, 46, 60, 67, show the
positions of $z \sim 1.61$ X-ray sources (see text).}
\label{fig:space_distr}
\end{figure}

The structure at $z ~\approx~ 1.61$ is extending across a transverse size of 
$\sim$ 5 Mpc in a wall-like pattern rather than a group structure
(see Fig.~\ref{fig:space_distr}).
\subsection{High redshift galaxies}
\begin{figure}
 \centering
 \includegraphics[width=\textwidth,height=12cm]{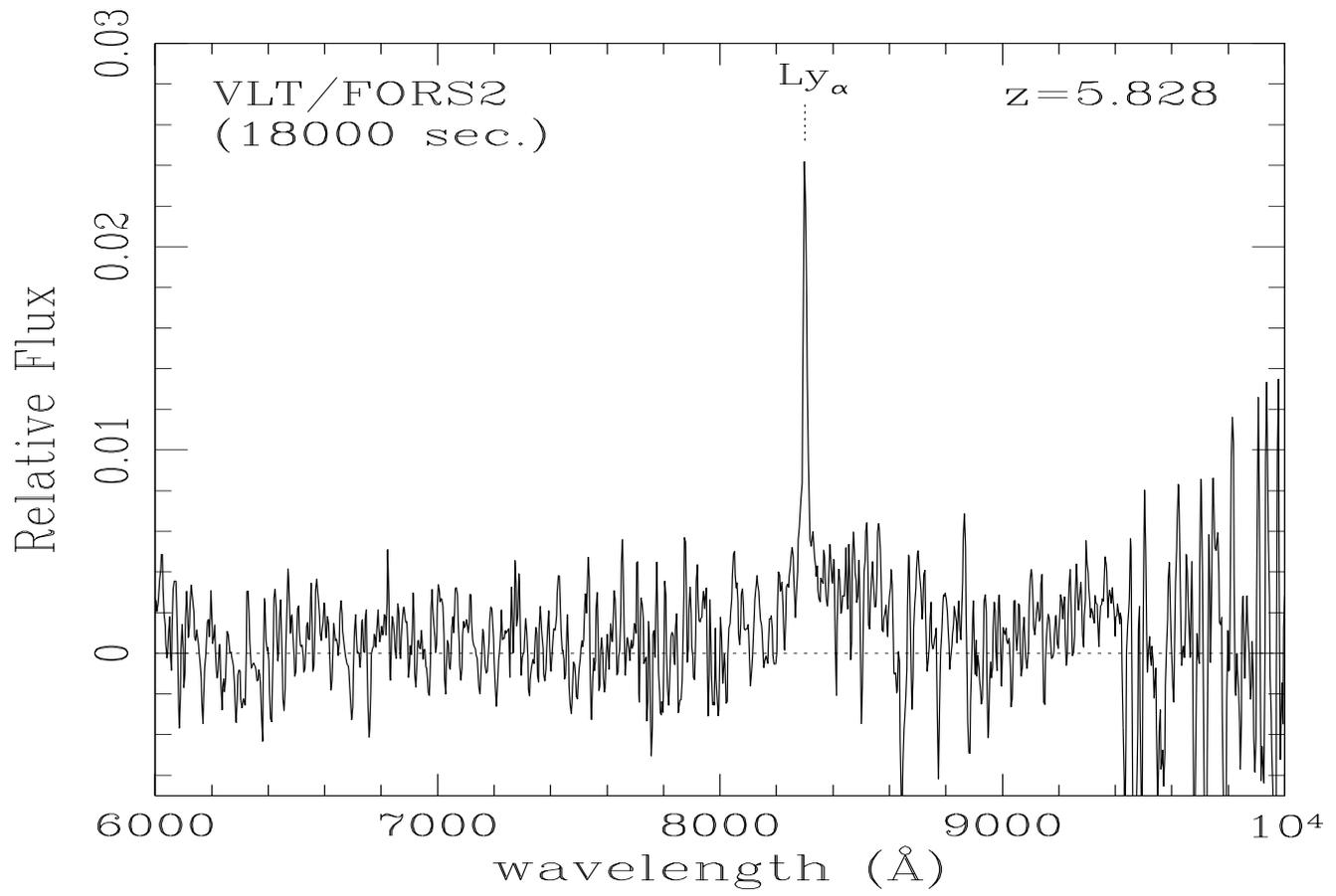}
 \caption{VLT spectrum of the $i_{775}$--dropout galaxy GDS~J033240.01-274815.0.}
\label{fig:high_z5.83}
\end{figure}
\begin{figure}
 \centering
 \includegraphics[width=\textwidth]{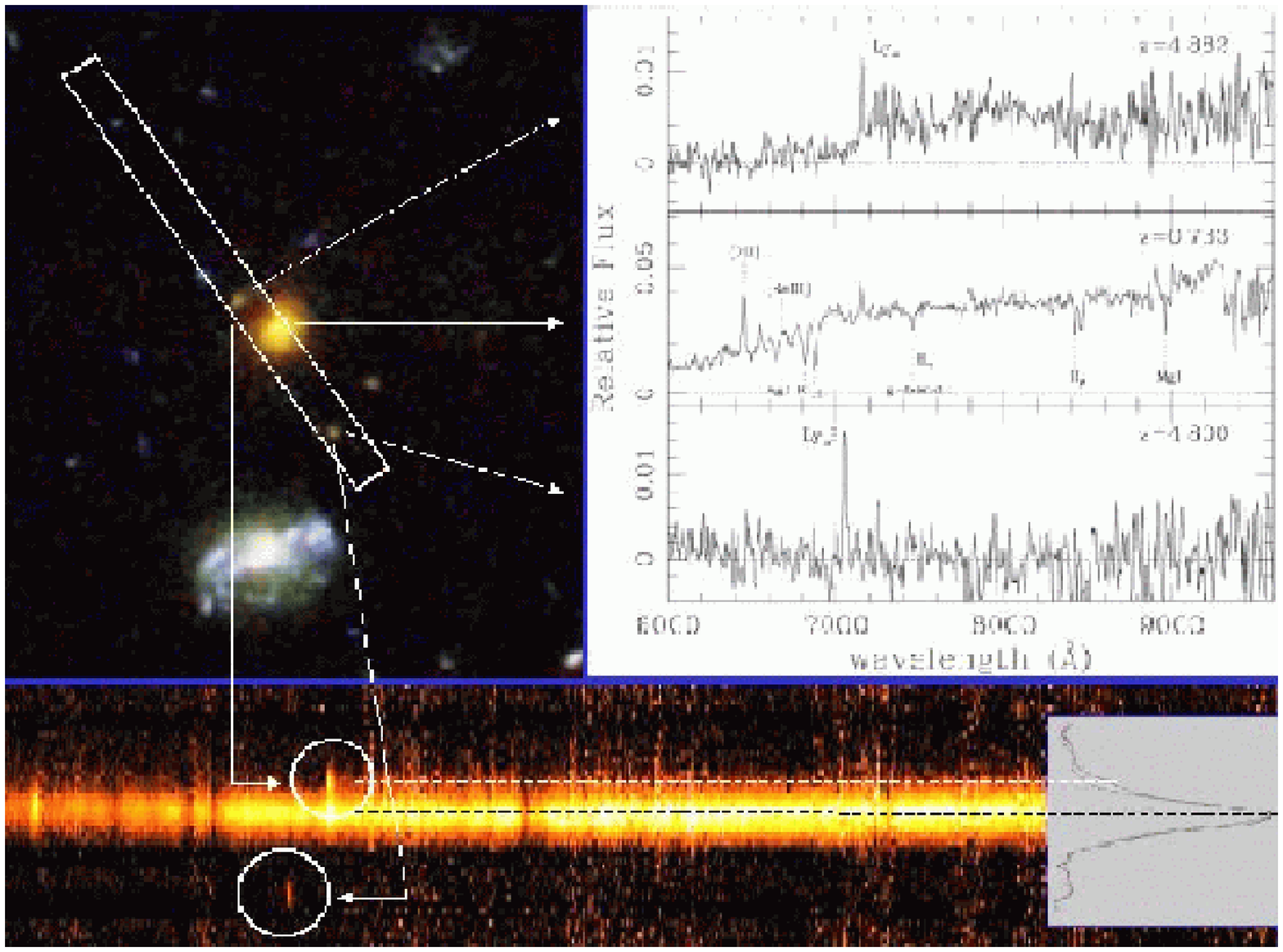}
 \caption{Simultaneous spectrum of three sources in the slit. On the
 right of the figure, the 1D spectra of the z=0.733 main galaxy
 $GDS~J033228.88-274129.3$, the single emission line $\sim 3$ arcsecond
 below ($GDS~J033228.84-274132.7$) and the object $\sim 1.5$ arcsecond 
 above  are shown. The left-hand panel shows the ACS color image,
 5 arcsec on a side. North is up, east is to the left. The bottom panel
 shows the 2D spectrum, with the spatial profile obtained by collapsing
 80 columns (256 \AA), centered at 7150\AA, shown to the right.
 Candidate serendipitous $Ly_{\alpha}$ emission lines are clearly marked.
 The object above the target source shows faint continuum reward of the 
 emission line.}
\label{fig:peculiar1}
\end{figure}
As discussed in Sect. 2, the target selection includes mainly low
redshift objects ($z < 2$).
For three galaxies, however, a redshift larger than four was measured:
the galaxy $GDS~J033240.01-274815.0$ at $z=5.828$ the only $i_{775}$-dropout
(see Sect. 2) actually targeted in the present observations 
and two serendipitously-observed high redshift sources, 
$GDS~J033228.84-274132.7$ and one object
at $\alpha = 3^h \ 32^m \ 28.94^s$, $\delta = -27^{\circ} \ 41' \ 28.19''$
not present in the catalog v1.0, measured at $z=4.800$ and $z=4.882$, 
respectively.

The $i$-dropout candidate has been observed with both the Keck and VLT
telescopes (\cite{dick04}).
In Figure~\ref{fig:high_z5.83} the FORS2 spectrum of the $i$-dropout
source is shown.
The $Ly_{\alpha}$ line is clearly detected at $z = 5.828$ and
shows the blue cut--off characteristic of high--redshift $Ly_{\alpha}$ 
emitters and the $Ly_{\alpha}$ forest continuum break.

Figure~\ref{fig:peculiar1} shows a peculiar system of three sources: 
two emission-line sources above ($\sim$1.5 arcsecond)
and below ($\sim$3 arcsecond) the main galaxy 
$GDS~J033228.88-274129.3$, clearly visible in the ACS color image
and in the two dimensional spectrum.
The same target has been observed in two different masks adopting the
same orientation of the slits. The total exposure time is
$\simeq$43 ks.
The extracted one dimensional spectra are shown in the right side of 
the Figure~\ref{fig:peculiar1}. 

The main galaxy $GDS~J033228.88-274129.3$ has a redshift $z=0.733$ 
with both emission and absorption lines measured (quality flag ``A''):
[O\,{\sc ii}]3727, MgI, Ca H and K, g-band, etc. 
The bottom object ($GDS~J033228.84-274132.7$) shows a
solo-emission line at 7052\AA\ ~(see the 1-D spectrum), 
and is not detected in the ACS B band, we interpret this line 
as $Ly_{\alpha}$ at $z=4.800$ with quality ``C''.

The source above $GDS~J033228.88-274129.3$ 
is most probably a $Ly_{\alpha}$ emitter at 
redshift $z=4.882$ (quality ``B''). The spectrum has been 
extracted subtracting the contamination of the tail
of the main galaxy. After the subtraction the shape of the 
spectrum shows the blue cut--off and the $Ly_{\alpha}$ forest continuum 
break, typical of the LBGs.

\subsection{Dynamical masses of galaxies at z $\sim$ 1}

\begin{figure}
 \centering
 \includegraphics[width=\textwidth]{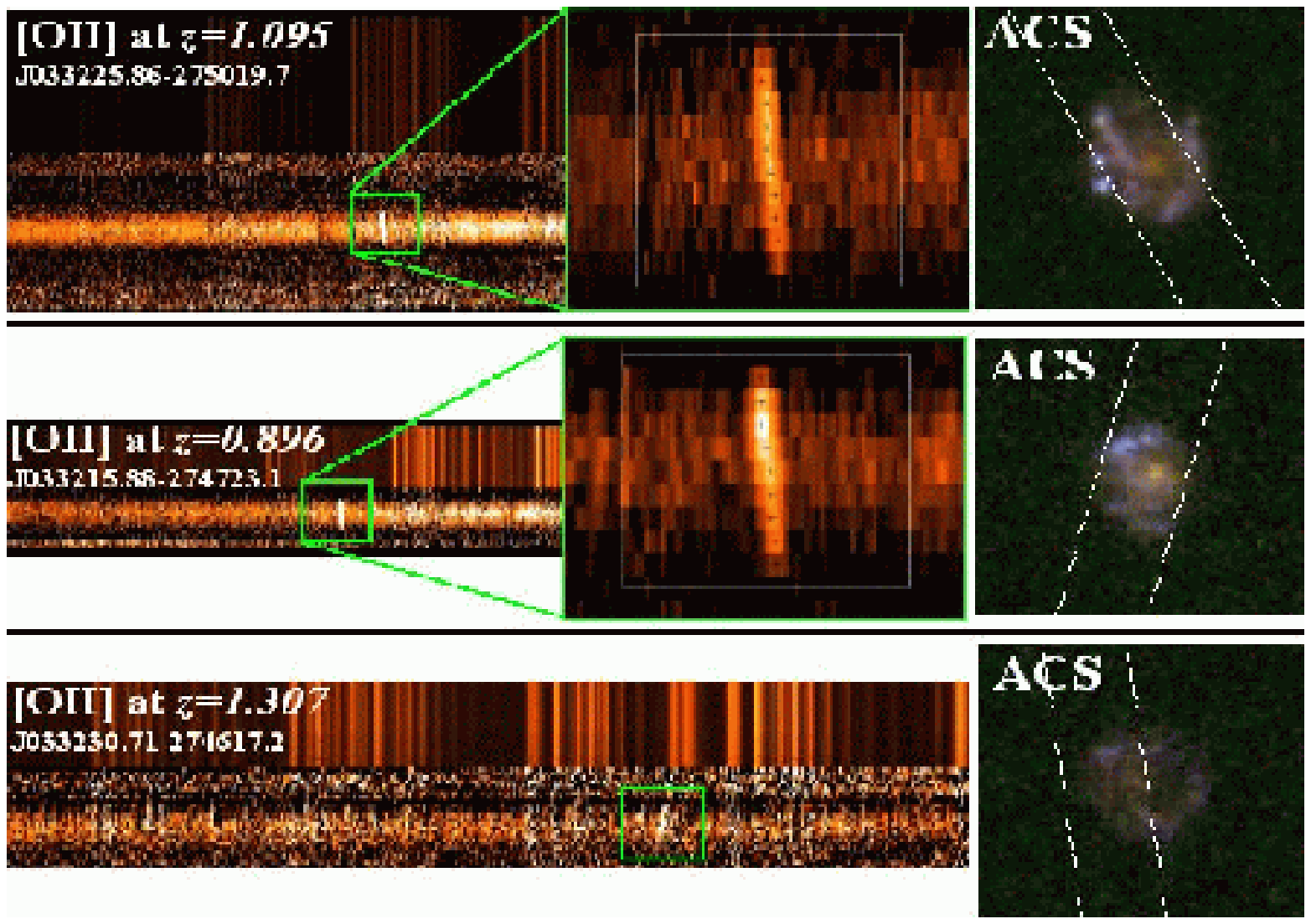}
 \caption{Three examples of tilted [O\,{\sc ii}]3727 emission line at redshift around 1. 
The two dimensional FORS2 spectra are shown (object and sky lines).
In the first two spectra (top and middle) a zoom of the [O\,{\sc ii}]3727 emission line 
is shown (the white rectangle underline the region where the Gaussian 
fit has been performed to derive the line peak, small black crosses),
in the bottom spectrum the line is too faint to calculate a reliable peak
(this object has been serendipitously-identified).
In the right side of the spectra the ACS images of the galaxies and the
slits orientation are shown.}
\label{fig:OIItilted}
\end{figure}

Three galaxies, $GDS~J033215.88-274723.1$, $GDS~J033225.86-275019.7$ and
$GDS~J033230.71-274617.2$, at redshift z=0.896, 1.095 and z=1.307
respectively show a spatially resolved [O\,{\sc ii}]3727 line with a
characteristic ``tilt'' indicative of a high rotation velocity (see
Figure~\ref{fig:OIItilted}).

Various studies have been carried out on the internal kinematics of
distant galaxies (\cite{vogt96,vogt97}, \cite{moor00}, \cite{pett01}
and \cite{DS01}).  \cite{rigop02} have determined velocity profiles
with a medium resolution grating R$\sim$5000 of three galaxies at
z$\sim$0.6 and one at z$\sim$0.8, detected by ISOCAM in the HDF--S.
For one object they have derived a rotational velocity of 460 km
s$^{-1}$ containing a mass of 10$^{12}M_\odot$ (within a radius of 20
Kpc) significantly higher than the dynamical masses measured in most
other local and high redshift spirals.

In the case of $GDS~J033215.88-274723.1$,~$GDS~J033225.86-275019.7$ and
$GDS~J033230.71-274617.2$, the spectra, in spite of the relatively low 
resolution $\Re\sim$860, clearly show a tilt of several pixels
(corresponding to about 10\AA). The measured velocity increases with 
increasing distance from the center
of the objects reaching a value of the order of and greater than 
400 km s$^{-1}$ at the extremes. 
For the object $GDS~J033225.86-275019.7$ we have measured a 
displacement between the two extreme peaks of 11.5\AA\, 
(top panel of the Figure~\ref{fig:OIItilted}), while a
displacement of 9.6\AA\ ~has been measured in the case of
$GDS~J033215.88-274723.1$ (middle panel of the Figure~\ref{fig:OIItilted})

Assuming that the observed velocity structure is due to dynamically-relaxed rotation,
then it is possible to
estimate the dynamical mass for the three galaxies shown in 
Figure~\ref{fig:OIItilted} (e.g. \cite{lequeux83}): $\frac{1.6}{sin^{2}(i)}\times$10$^{11}M_\odot$
for the galaxy $GDS~J033215.88-274723.1$ (within a radius of 7.8 Kpc)
and $\frac{3.1}{sin^{2}(i)}\times$10$^{11}M_\odot$ for the galaxy $GDS~J033225.86-275019.7$
(within a radius of 9.8 Kpc). 
The noisy spectrum of the galaxy $GDS~J033230.71-274617.2$ allows us to 
roughly measure a dynamical mass of the order of $\frac{1.5}{sin^{2}(i)}\times$10$^{11}M_\odot$
(within a radius of 7.5 Kpc).
The estimates should be considered a lower limit to the total 
dynamical mass because more external parts of the rotating structure
might have a lower surface brightness and remain undetected.

\subsection{$GDS~J033210.93-274721.5$: a spectrum contaminated by a nearby galaxy.}
\begin{figure}
 \centering
 \includegraphics[width=\textwidth]{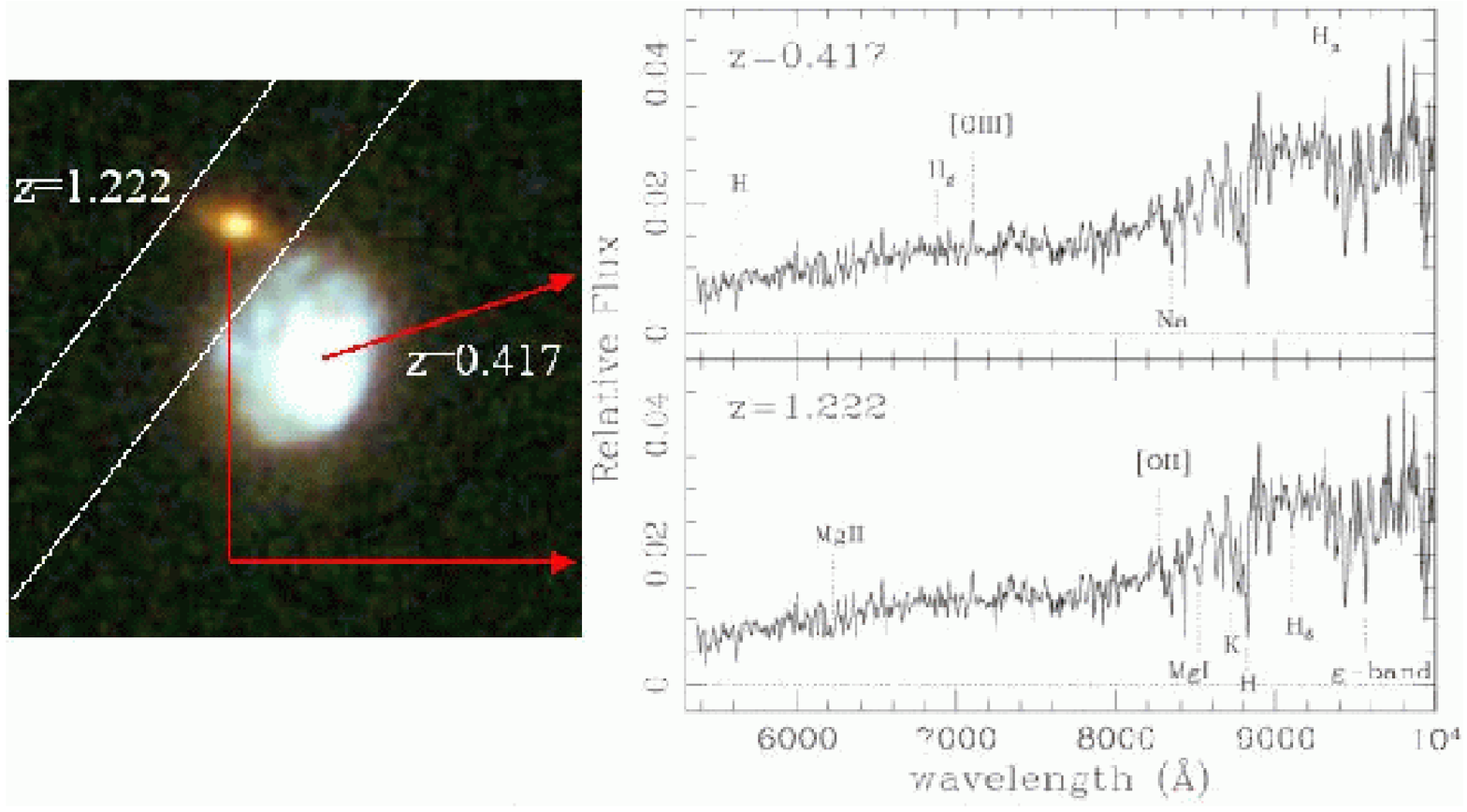} 
 \caption{The light merged case, two objects at different redshift superimposed
in the slit (marked with white lines in the left panel). In the right
panel the same extracted spectrum with different identifications.
An elliptical galaxy 
(the target, $GDS~J033210.93-274721.5$) at z=1.222 clearly identified with
the Ca H and K, H$\delta$, MgI (quality flag ``A''). 
The bright bluer object ($GDS~J033210.92-274722.8$) shows absorption and emission 
lines: Ca H, [O\,{\sc iii}]5007, Na, H$\alpha$ at $z=0.417$ 
(quality flag ``B''). The Ca K is contaminated by the sky line $\sim$5577\AA.}

\label{fig:light_merged}
\end{figure}

The spectrum of the galaxy $GDS~J033210.93-274721.5$ simultaneously
shows features corresponding to the redshifts z=1.222 and z=0.417 
(Figure~\ref{fig:light_merged}).  
The origin of the overlap is the presence of a nearby galaxy
($z_{850}=19.98$, $GDS~J033210.92-274722.8$) offset by 1.3 arcsecond 
with a redshift $z=0.417$. Light from the brighter $z=0.417$ galaxy
contaminates the spectrum of the fainter ($z_{850}=22.19$), 
higher redshift galaxy $GDS~J033210.93-274721.5$ (see
Figure~\ref{fig:light_merged}). 
Such cases may represent a problem and a source of error
in large spectroscopic surveys, which require an highly automated
data processing. A possible solution is to evaluate a priori on
the basis of imaging what are the cases subject of light contamination
requiring a ``special'' reduction. Alternatively, color-redshift
diagrams (such as Figure~\ref{fig:i_zVSzspec}), a comparison of
spectroscopic and photometric redshifts or similar diagnostics are
required to carry out the necessary data quality control and identify
possible misidentifications.

\section{Conclusions}
In the framework of the Great Observatories Origins Deep Survey a
large sample of galaxies in the Chandra Deep Field South has been
spectroscopically targeted.
A total of 303 objects with $z_{850} \mincir 25.5$ has been observed
with the FORS2 spectrograph at the ESO VLT providing $234$ redshift
determinations. 
From a variety of diagnostics the measurement of the redshifts appears
to be highly accurate (with a typical $\sigma_z = 0.001$) and reliable
(with an estimated rate of catastrophic misidentifications at most few percent).
The reduced spectra and the derived redshifts are released to the community 
($\it{http://www.eso.org/science/goods/}$).
They constitute an essential contribution to reach the scientific goals
of GOODS, providing the time coordinate needed 
to delineate the evolution of galaxy masses, morphologies, and star
formation, calibrating the photometric redshifts that can be derived from the
imaging data at 0.36-8$\mu$m and enabling detailed studies 
of the physical diagnostics for galaxies in the GOODS field.

\begin{acknowledgements}
 We are grateful to the ESO staff in Paranal and Garching who greatly helped
 in the development of this programme.
 The work of DS was carried out at the Jet Propulsion Laboratory,
 California Institute of Technology, under a contract with NASA.
 L.A.M. acknowledges support by NASA through contract number 1224666 
 issued by the Jet Propulsion Laboratory, California Institute of Technology 
 under NASA contract 1407. We thank the ASI grant I/R/088/02 (SC, MN, EV).
\end{acknowledgements}

\end{document}